\documentclass[apj]{emulateapj}
\usepackage{apjfonts}
\usepackage{multirow}
\usepackage{color}
\usepackage{natbib}
\newcommand{\beq}{\begin{equation}}
\newcommand{\eeq}{\end{equation}}
\newcommand{\beqa}{\begin{eqnarray}}
\newcommand{\eeqa}{\end{eqnarray}}

\def\re{r$_\mathrm{e}$}

\shorttitle{Spectral Energy Distribution Mapping}
\shortauthors{Amblard et al.}

\begin{document}
\title{Spectral Energy Distribution Mapping of Two Elliptical Galaxies on sub-kpc scales}

\author{
A. Amblard\altaffilmark{1,2},
P. Temi\altaffilmark{1}, 
M. Gaspari\altaffilmark{3,4},
F. Brighenti\altaffilmark{5}
}

\altaffiltext{1}{NASA Ames Research Center, Moffett Field, CA, USA}
\altaffiltext{2}{BAER Institute, Sonoma, CA, USA}
\altaffiltext{3}{Department of Astrophysical Sciences, Princeton University, Princeton, NJ 08544, USA}
\altaffiltext{4}{{\it Einstein} and {\it Spitzer} Fellow}
\altaffiltext{5}{Astronomy Department, University of Bologna, Via Ranzani 1, 40127, Bologna, Italy}

\begin{abstract}
We use high-resolution  {\it Herschel}-PACS data of 2 nearby elliptical galaxies, IC1459 
\& NGC2768 to characterize their dust and stellar content. IC1459 \& NGC2768 have an unusually 
large amount of dust for elliptical galaxies (1$\,$to$\,$3$\times$10$^5$ M$_\odot$), this dust is also not 
distributed along the stellar content. Using data from GALEX (ultra-violet) to PACS (far-infrared), 
we analyze the spectral energy distribution (SED) of these galaxies with CIGALEMC as a function of 
the projected position, binning images in 7.2'' pixels. From this analysis, we derive maps of SED 
parameters, such as the metallicity, the stellar mass, the fraction of young star and the dust mass. 
The larger amount of dust in FIR maps seems related in our model to a 
larger fraction of young stars which can reach up to 4\% in the dustier area. The young stellar population 
is fitted as a recent ($\sim$ 0.5 Gyr) short burst of star formation for both galaxies. The metallicities, 
which are fairly large at the center of both galaxies, decrease with the radial distance with fairly steep 
gradient for elliptical galaxies. 

\end{abstract}

\section{Introduction}
Early-type galaxies (ETGs) have long been thought to be depleted of any molecular or atomic gas and to 
produce a negligible amount of stars. 
However, recent advances in technologies have allowed us to probe further the ISM of early-type galaxies 
and to realize that a few ETGs harbor a substantial 
amount of gas \citep{wiklind89,lees92,wang92,wiklind95,young11,alatalo13}.
ETGs generally have a complex multiphase interstellar medium (ISM), which is a key component of the
galactic ecosystem. Hot, warm and cold gas detections have been reported 
by several authors \citep[eg.][]{caon00,mathews03,sarzi06,sarzi13,mulchaey10,ellis06,davis11,serra12}
The cold ISM is often spatially extended, with irregular distribution and kinematics \citep{caon00}.
Dust is also present in all the phases of the ISM \citep{temi05,temi09,smith12}, and the dust-to-gas
ratio can be used to track down the origin of the cold gas.
A multiphase ISM is expected to keep the galaxy ``alive'', allowing for (perhaps tiny) episodes 
of star formation (SF) and a sizable fraction of ETGs shows evidence of recent SF \citep{trager00,kaviraj07}.
In some ETGs a fraction of cold gas is thought to have an external origin, as indicated by 
the misaligned kinematics of the cold gas (both ionized and neutral) with respect to the stars 
\citep[e.g.][and references therein]{davis11}.
However, some ISM in the cold phase certainly comes from internal processes, such as stellar mass 
loss or hot gas cooling \citep{davis11,werner14,david14}.\\
ETGs are viewed as the ending step of galactic evolution, as such
their proportion is increasing as the Universe is aging and they
contain important clues of how star formation evolved in galaxies. The proximity
of some of these nearby objects is the perfect testing ground for theories, given the 
accessibility to high angular resolution and spectroscopy that some higher redshift 
objects lack.\\
Numerical simulations show that ETGs can be produced from major mergers of late-type galaxies. 
During mergers, tidal interactions drive the gas into the center, this gas fuels a starburst 
and feeds the black hole (BH) growth. The gas consumption by the starburst and the BH feedback
leave a merger remnant with a very low star formation rate \citep{dimatteo05,springel05a,springel05b,hopkins09}.
However the merger rate needed to produce the correct amount of ETGs is larger than the one estimated with the 
same simulations.\\
A first attempt to  distinguish gas-rich and gas-poor ETGs split them in two separate groups, 
lenticular and ellipticals. 
Morphologically, lenticular galaxies still have a fairly large disk that does not present spiral arms, whereas 
elliptical galaxies are bulge-dominated.
More recently, ATLAS-3D survey has tried to separate more accurately two sub-class of ETGs, using the gas dynamics
\citep{emsellem11}. 
They separate their ETGs sample into a slow-rotator category that roughly corresponds to giant ellipticals and a 
fast-rotator category that is composed mostly of lenticulars.\\
Lenticular galaxies seems to have a wider range of properties compared to ellipticals that resemble more 
the old definition of ETGs. However, even in ellipticals, there subsists large differences.
Recent observations of elliptical galaxies with {\it Spitzer} and Herschel
\citep{temi05,temi07a,temi07b,temi09,mathews13,smith12,agius13}
have revealed that the far-infrared luminosity $L_{FIR}$ from these
galaxies can vary by $\sim$100 among galaxies with similar optical luminosity.
The 70$\mu$m band luminosities \citep[from][]{temi07a,temi09},
is a good example of such a huge scatter in the far-IR luminosity of elliptical galaxies.
Some of the high $L_{70}$ galaxies, are members of a small subset of ellipticals
having radio detections of neutral and molecular gas.
Few others may be S0 galaxies which, because of their rotationally-supported disks, often
contain large masses of cold gas and dust.
Ellipticals containing large excess masses of dust and cold gas probably results from significant galaxy 
mergers in the past.
However a fraction of elliptical galaxies appear to be completely {\it normal} but $L_{70}$ in these 
galaxies still ranges over a factor of $\sim30$, far larger than can be explained by
uncertainties in the estimate of the far-IR SED  due to local stellar mass loss.
While a significant fraction of the cold gas mass in low to intermediate mass ETGs is thought
to have an external, merger-related origin \citep[e.g.][]{davis11},
 in the most massive ETGs the cold gas phases are presumably generated internally 
\citep{davis11,david14,werner14}.
Mergers with gas and dust-rich galaxies have often been suggested for the origin of dust in all elliptical galaxies
\citep[e.g.][]{forbes91}.
Although the merger explanation is almost certainly correct in some cases, 
mergers cannot explain most of the observed scatter in $L_{70}$.
A crucial element in our understanding of the evolution of galaxies toward ETGs is the mutual 
role played by the major merging of galaxies and the secular star-formation quenching. 
Neither of these scenarios account yet for all the observational evidences, and one could assume both 
contributing to some extent. \\
This paper concentrates on the study of 2 atypical elliptical galaxies, that contains a
substantial amount of dust, IC1459 \& NGC2768.
Both galaxies have optical and near-IR morphologies that resemble any large elliptical galaxies, they both 
are fairly red. IC1459 has a B-V and V-R color of 0.9 according to \cite{forbes95} and 
a g-r color of 0.85 in the OmegaCAM survey, NGC2768 has a g-r color of 0.87 in SDSS.
However a first hint of their peculiarity came from dust absorption images that
show unusual dust features \citep{malin85,forbes95,kim89}. 
Furthermore, observations of dust emission 
in the far-infrared 
revealed some larger than expected fluxes and unusual morphology \citep{temi07a,crocker08}.\\
In this paper, we utilize the recent data from Herschel PACS photometer at 70 and 160 $\mu$m, that reached the
unprecedented FIR high angular resolution of 6'' and 12'' respectively to map the parameters from a stellar 
population synthesis (SPS) model with wide-band photometric observations from the UV to the FIR. 

\section{Galaxies sampled}

\subsection{IC 1459}
IC1459 is a massive elliptical galaxy E3 (M$_V$ = -21.19, D = 16.5 h$^{-1}$ Mpc) that belongs to a group of mostly spiral galaxies \citep[group number 15 of][]{huchra82}.
It has a counter-rotating stellar core \citep{franx88} with a maximum rotation of about 170 km s$^{-1}$ at 2'' (0.15 kpc). 
The stellar mass of the counter-rotating core is about 10$^{10}$ M$_\odot$ \citep{franx88} and 
the outer region velocity goes up to 45 km/s but in opposite direction.
Stellar shells at tens of kpc from the center and other peculiar isophotes were observed by \cite{williams79,malin85,forbes95} and quantified by \cite{tal09}.
This disturbed morphology has led to the hypothesis that IC1459 accreted some external material in the past,
probably from a merger \citep{cappellari02}.\\
IC1459 hosts an active galactic nucleus characterized by two strong symmetric radio jets \citep[about 1 Jy at 1.4 GHz;][]{slee94,ekers89,tingay15}, whose activity may have been triggered by the same event that gave the galaxy its peculiar morphology and kinematics. 
IC1459 has a strong LINER optical spectrum \citep{phillips86}.
However Chandra observations of the supermassive black hole (SMBH) of IC 1459 show a weak X-ray source \citep[L$_X$ = 8$\times$10$^{40}$ ergs.s$^{-1}$, 0.3-8 keV,][]{fabbiano03} with a flux much lower than expected from a 
 2$\times$10$^9$ M$_\odot$ black hole \citep[10x lower than normal radio loud galaxy, ][]{cappellari02}.
IC1459 has also been found to be a Gigahertz-Peaked Spectrum (GPS) radio source \citep{tingay03}
with a peak frequency of 2.5 GHz and low power.\\
\cite{serra10} found that IC1459 has a SSP(single stellar population)-equivalent age 
of 3.5$^{+1.7}_{-0.4}$ Gyr using spectral line data from \cite{tal09}
and SSP models from \cite{thomas03}. This age estimate is biased towards the age of the youngest population 
\citep{serra07} and is more an indication of the age of the young stars. Serra et al. 2010 gave
a rough estimate of 0.5 to 5 \% of young stellar population (in mass) formed between 300 Myr and 1 Gyr ago.
\cite{prandoni12} deduced a molecular gas mass of about 2.5$\times 10^7$M$_\odot$ using CO(2-1) observations
taken with the Atacama Pathfinder Experiment telescope (APEX), while \cite{serra10} measured a HI mass of about
2.5$\times 10^8$M$_\odot$.

\subsection{NGC 2768}
NGC2768 has been classified as an E6 galaxy in the third reference catalogue of bright galaxies 
\citep[RCC3,][]{devaucouleurs91}, but \cite{sandage94} have reported the galaxy
as an almost edge-on S0 galaxy in the Carnegie Atlas of Galaxies. The galaxy is part of the SAURON sample \citep{emsellem07}, where it is classified as a fast-rotator 
with a single component in their stellar kinematic classification scheme. 

NGC 2768 has a low-luminosity active galactic nucleus (AGN) with a low-ionisation nuclear emission-line region (LINER) spectrum \citep{heckman80}, a compact radio core \citep{nagar05} and an X-ray source consistent with being a point source \citep{komossa99}, its X-ray luminosity is about 10$^{40}$ erg/s \citep{boroson11}.

It is a fairly isolated galaxy \citep{wiklind95} but has been associated with the Lyon Group of
 Galaxies \citep{garcia93}.
The effective radius of the galaxy is about 67'' \citep{devaucouleurs91}, corresponding to 7.3 kpc 
at a distance of 22.4 Mpc \citep{tonry01}. \cite{kim89} discovered a dust lane along the minor axis of 
NGC2768, measured the rotation of the ionized gas around the major axis, and suggested an external 
origin of the gas in this galaxy.  \cite{crocker08} traced the molecular gas of NGC 2768 with the 
CO(1-0) and (2-1) line emission, finding a molecular polar disc, coincidental with the dust absorption
of \cite{kim89}. \cite{osullivan15} inferred a molecular gas mass of 2$\times 10^7$M$_\odot$ and \cite{morganti06}
measured a HI mass of about  2$\times 10^8$M$_\odot$.

\section{Data}
\label{sec:dat}

We base our work primarily on data obtained with {\it Herschel} space telescope \citep{pilbratt10}
using the PACS instrument with the blue (70 $\mu$m) and red (160 $\mu$m) filters.
In order to gather a larger amount of information in a self-consistent manner from the SED-fitting, 
 we collect data from 6 additional instruments which span wavelengths from the UV to the FIR and 
have extensive sky coverage : GALEX for the UV part, SDSS \& OmegaCAM for the optical, 2MASS and IRAC/MIPS-Spitzer
 for the Near-InfraRed (NIR) and Mid-Infrared (MIR).

\subsection{UV data}

To cover the UV part of the spectrum we used GALEX GR6 data release\footnote{http://galex.stsci.edu/GR6/}. GALEX surveys cover 25,000 deg$^2$ of the sky with a sensitivity down to m$_{AB}$=21 for the All Sky Imaging Survey (AIS) and m$_{AB}$=25 for the Deep Imaging Survey \citep[DIS; ][]{morrissey07}. GALEX is a NASA satellite, equipped with two microchannel plate detectors imaging in the near-UV (NUV) at 2271 {\AA}  and far-UV (FUV) at 1528 {\AA}  and a grism to disperse light for low resolution spectroscopy.  The source position accuracy is about 0.34 arcseconds and the angular resolution of FUV and NUV is respectively 4.2 and 4.9 arcseconds. We applied a galactic dust extinction correction, A(FUV)/E(B-V)=8.376 A(NUV)/E(B-V)=8.741, to GALEX data, assuming Milky Way dust with R$_v$=3.1 \citep{cardelli89,marino11}. UV emission is a good indicator of the dust content and star formation rate of galaxies when compared with optical data. NGC2768 is a galaxy of the nearby galaxy survey that has a depth $\mu_{AB}$ of 28 mag/sq. arcsec., IC1459 was observed by GALEX with the Guest Program Cycle 1 down
to a similar depth. Both have been detected in the FUV and NUV.

\subsection{Optical, NIR and MIR Observations}

SDSS 5 bands cover the optical range of the SED and were observed on a large portion of
the sky (14,555 deg$^2$). The SDSS data have an angular resolution
of about 1.5 arcseconds.  We retrieved SDSS data through the Imaging Query Form interface\footnote{http://skyserver.sdss3.org/dr9/en/tools/search/IQS.asp}. N2768 has been observed in the 5 bands of the Sloan Digital Sky Survey  (SDSS), u, g, r, i and z (respectively 0.335, 0.469, 0.616, 0.748 and 0.893$\,\mu m$).
Unfortunately, IC1459 is too south to have been observed by SDSS, for this galaxy we used the
data from the OmegaCAM science archive\footnote{http://osa.roe.ac.uk/}. This archive
contains data obtained on the VLT Survey Telescope (VST) mostly from the VST ATLAS survey \citep{shanks15}.
The VST ATLAS survey covers 4,500 deg$^2$ in the southern hemisphere at high galactic latitude
with depth comparable to SDSS and the same set of 5 filters.
This archive contains images of IC1459 in 4 filters g, r, i and z with an angular resolution of about 0.8 to 1.0 arcseconds. We applied a galactic dust extinction correction with the values provided on the 
NASA/IPAC Extragalactic Database (NED)\footnote{http://ned.ipac.caltech.edu/}.
\\
At near-infrared wavelength, we use the extended source catalog of the Two Micron All Sky Survey (2MASS) data, which contains 1,647,599 sources. 2MASS resolution is about 2 arcseconds and its source position accuracy
is about 0.5 arcseconds. The 10-$\sigma$ limiting magnitude in the 3 filters J,H,K$_s$ is about 14.7, 13.9, 13.1
\footnote{http://www.ipac.caltech.edu/2mass/releases/allsky/doc/sec4\_5.html}.
Both galaxies have counterparts in the 3 different filters, J, H and K$_s$ bands, 
at 1.24, 1.66 and 2.16$\,\mu m$ respectively. \\
The {\it Spitzer} Space Telescope provides data in the near-IR with the IRAC camera with 
4 channels imaging at 3.6, 4.5, 5.6 and 8$\,\mu m$ with about 2 arcsecond angular resolution 
and in the mid-IR with the MIPS instrument observing at 24 \,$\mu m$ (at longer wavelengths
we used {\it Herschel} PACS) at an angular resolution of 6 arcseconds \citep{rieke04}. 
We download data for our galaxies from the NASA/IPAC Infrared Science Archive\footnote{http://sha.ipac.caltech.edu/applications/Spitzer/SHA}.

\subsection{FIR Observations}

The launch of the {\it Herschel}\footnote{Herschel is an ESA space observatory with science 
instruments provided by European-led Principal Investigator consortia and with important
 participation from NASA.} telescope allowed unprecedented precisions at FIR wavelength.
We used public level2 data from the PACS instrument, downloaded from the Herschel Science 
Archive\footnote{http://herschel.esac.esa.int/Science\_Archive.shtml}. The PACS instrument observed
at 70 and 160$\,\mu$m  (the 100$\,\mu$m was not available for our sources) with an angular
resolution of about 6 and 12 arcseconds \citep{poglitsch10}.
Level2 maps were combined into a single map for each objects using a simple pixel co-addition
technique. We subtract to each image a background, estimated by taking a median at 3 arcminutes around the source.

\section{SED Analysis}

\subsection{Data pre-processing}
In order to compare images taken by different instruments, each map was converted in units of Jy/pixel
with a common angular resolution of 12'' and aligned so that they would cover the same area of the sky. 
The original angular resolution of each dataset was taken 
as 4.2'' and 4.9'' for GALEX \footnote{http://www.galex.caltech.edu/researcher/techdoc-ch5.html}
, 1.4'' for SDSS (median in r\footnote{http://www.sdss.org/dr3/}), between 0.8'' and 1.0''
for OmegaCAM data \citep{shanks15}, about 2.8'' for 2MASS
\footnote{http://www.sao.arizona.edu/FLWO/pairitel/seesum.html}
1.66'', 1.72'', 1.88'', 1.98'' for IRAC \footnote{http://irsa.ipac.caltech.edu/data/SPITZER/docs/irac}, 6'' for 24$\,\mu$m MIPS \footnote{http://irsa.ipac.caltech.edu/data/SPITZER/docs/mips}, 5.8'' and 12.0'' for PACS \citep{poglitsch10}. For each wavelength, a convolution kernel was computed to bring the PSF to 12'' and each map was 
convolved using the appropriate kernel. To fit the SEDs of individual area, images were pixelated to 7.2'' to sample adequately the 12'' FWHM beam. Consequently neighbor pixels are slightly correlated.

\subsection{SPS model}

To fit the spectral energy distribution (SED) of our galaxies, we use CIGALEMC\footnote{http://cigale.oamp.fr/} 
\citep{serra11,amblard14} which is a modified version of the Code Investigating GALaxy Emission 
\citep[CIGALE, ][]{noll09,giovannoli11}. CIGALEMC uses a Markov Chain Monte Carlo sampling of the 
CIGALE parameters which allows to increase the size of the parameter space covered and a more efficient sampling of it.
CIGALEMC uses the \cite{maraston05} stellar population model and we use the Salpeter initial mass function (IMF) 
\citep{salpeter55}.\\ 
The Salpeter IMF is in general a better match for massive early-type galaxies as has been 
found previously \citep{grillo09,auger10,treu10,spiniello11} using stellar dynamics and gravitational lensing. 
The \cite{maraston05} stellar population model includes a realistic treatment of the thermally pulsating asymptotic
 giant branch (TP-AGB). The TP-AGB phase model is important to derive an accurate stellar mass 
\citep{maraston06,ilbert10}. The metallicity $Z$ is fitted with an uniform prior between 0.005 and 0.07, 
a previous attempt to fit our SEDs with a fixed solar metallicity did not return satisfactory fits.\\
When fitting the data, we assume an exponentially decreasing star formation rate (SFR) for the old  
star population following \cite{giovannoli11}. The age of the old stellar population is constrained between 8 and 14 Gyr and the e-folding time of the old population is constrained between 0.5 and 3 Gyr.
The star formation history of the young stellar population is assumed to be Gaussian-shaped \citep{smith15}.
The peak age is constrained to be between 0.1 and 4 Gyr and the width ($\sigma$) between 0.02 and 1.5 Gyr.\\

We use the \cite{calzetti94} and \cite{calzetti97} attenuation to describe the dust 
absorption of star light. We do not add any modification to the Calzetti curve, like a 2175 {\AA} UV bump or a 
change of slope. The attenuation is modeled independently for the old and young stellar populations, the attenuation 
factor for the young population is A$_{\mathrm{V}}$ (V band attenuation) and there is a reduction factor f$_V$ for the old 
stellar population (A$_{\mathrm{V}} \times$ f$_V$). For simplicity, here we assumed that both the old and young population have the same extinction and fix f$_\mathrm{V}$ to 1.
\\
The IR emission from the dust is computed using \cite{dale02} model, which is composed of 64 templates parametrized 
by a slope $\alpha$. This slope represents the power-law slope of the dust mass over the heating intensity.
 \cite{dale02} followed \cite{desert90} approach by dividing their dust emission sources into large grains, small 
grains and PAHs. They normalized these components using observations from IRAS, ISO and SCUBA. In order to retrieve from $\alpha$, some commonly used quantities such as a dust temperature and a $\beta$ index, we fit a relation between $\alpha$ and $\beta$ and $T_\mathrm{d}$. 
This implies that our measurements of  $\beta$ and $T_\mathrm{d}$ are not fully independent being linked by the \cite{dale02} templates.\\
 CIGALEMC also includes a model for the AGN emission, using the AGN templates from \cite{siebenmorgen04a,siebenmorgen04b} and a parameter 
for its amplitude f$_{AGN}$.
The ten free parameters of the fit, {Z, $\tau_{old}$, t$_{old}$, $\sigma_{young}$, t$_{young}$, f$_{young}$, A$_{\mathrm{V}}$, $\alpha$, 
f$_{AGN}$, M$_{gal}$, are described in table \ref{tab:1} along with their priors. 

\begin{deluxetable}{lll}[ht!]
\tablecaption{Parameters fitted by CIGALEMC to the galaxy SEDs, \\with priors chosen.}
\tablehead{\colhead{Parameters} & \colhead{Priors} & \colhead{Description}}
\startdata
Z & 0.005 < Z < 0.07 & metallicity\\
t$_{old}$ &   8.0  <$t_{old}$< 14 Gyr & old star population age\\
$\tau_{old}$  &  0.5 <$\tau_{old}$< 3 Gyr & old star population e-folding time\\ 
t$_{young}$ & 0.1 <t$_{young}$< 4 Gyr & age of the young stellar population\\
$\sigma_{young}$ & 0.02 < $\sigma_{young}$ < 1.5 Gyr & time-spread of young stars\\
f$_{young}$ & 0 <f$_{young}$< 1 & fraction of young stars\\
A$_{\mathrm{V}}$ & 0.01 <A$_{\mathrm{V}}$< 5 mag. & dust extinction in the V band\\
$\alpha$ & 0.2 <$\alpha$< 3.9 & slope of the dust mass over heating\\
f$_{AGN}$ & 0 <f$_{AGN}$< 1 & AGN fraction of the dust luminosity\\
M$_{gal}$ & 2 <M$_{gal}$< 11 & logarithm of the galaxy mass
\enddata
\label{tab:1}
\end{deluxetable}

In the analysis of the SEDs, we use some derived parameters : SFR, M$_*$ (the stellar mass) and M$_{d}$ (dust mass); 
these are computed either from the fitted parameters and/or from the fitted SED.
The SFR is computed from the contribution of the young and old stellar populations, however the young stellar population
 gives in general the dominant contribution. 
Therefore the SFR is mostly depending on the fitted parameters: 
the normalization M$_{gal}$, the fraction of young star f$_{young}$, and the age of young star population $t_{young}$. 
SFR increases with M$_{gal}$ and f$_{young}$, it decreases slightly as young stellar population ages, i.e. 
as $t_{young}$ gets larger. Apart from its M$_{gal}$ dependency, the SFR is constrained by the UV, optical 
and NIR part of the spectrum via the stellar population synthesis (SPS) of \cite{maraston05}.\\ 
M$_*$ is calculated by integrating over the evolution track of the \cite{maraston05} model, 
and depends primarily on the UV, optical and NIR part of the spectrum, except for the overall normalization 
defined by M$_{gal}$. M$_{d}$ is calculated from the dust parameter $\alpha$, the dust absorption parameter A$_{\mathrm{V}}$ and the mass of the galaxy M$_{gal}$.\\

\begin{figure*}[h!t]
\begin{center}
\includegraphics[width=8cm]{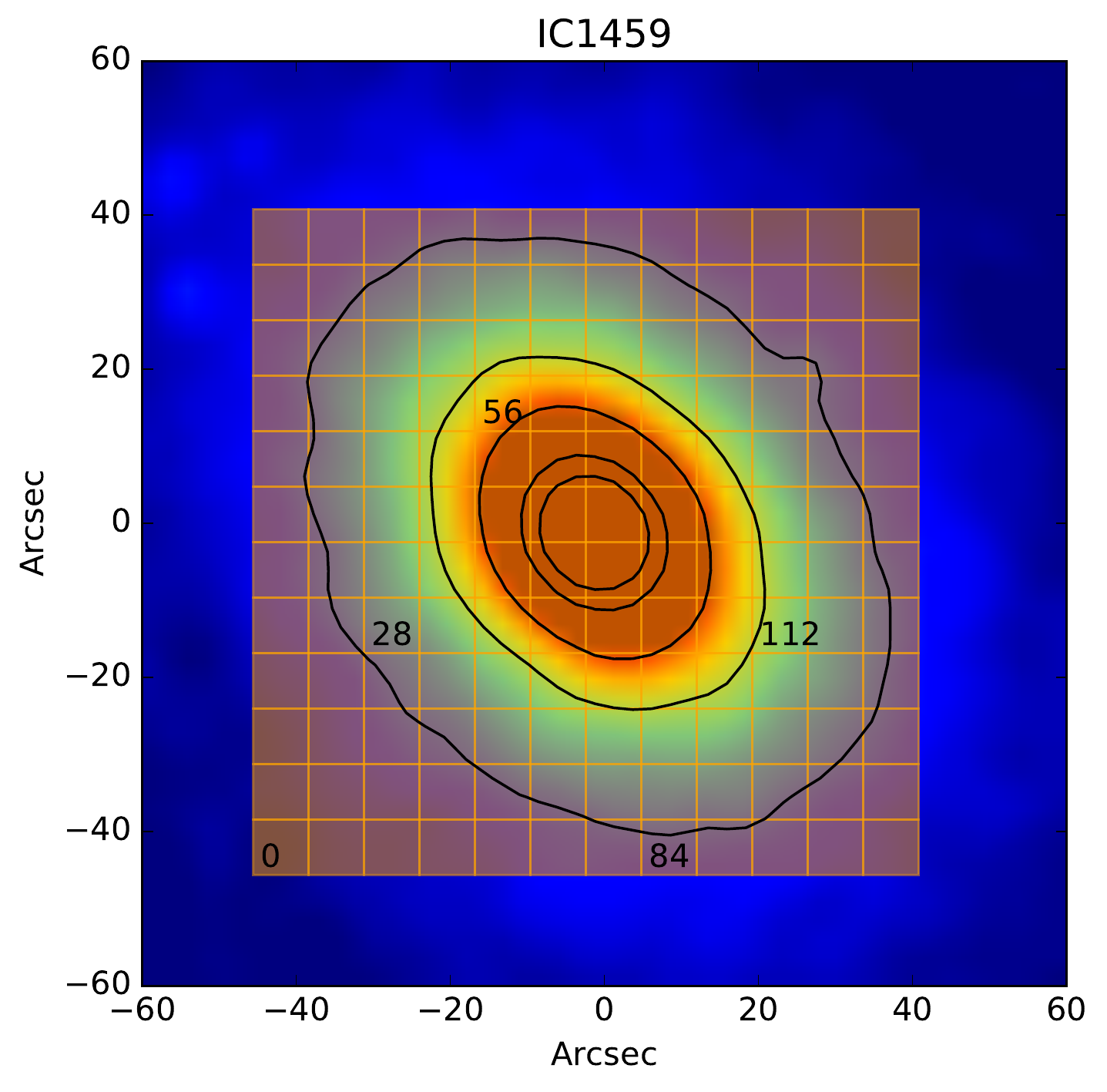}\hspace{0.5cm}\includegraphics[width=8cm]{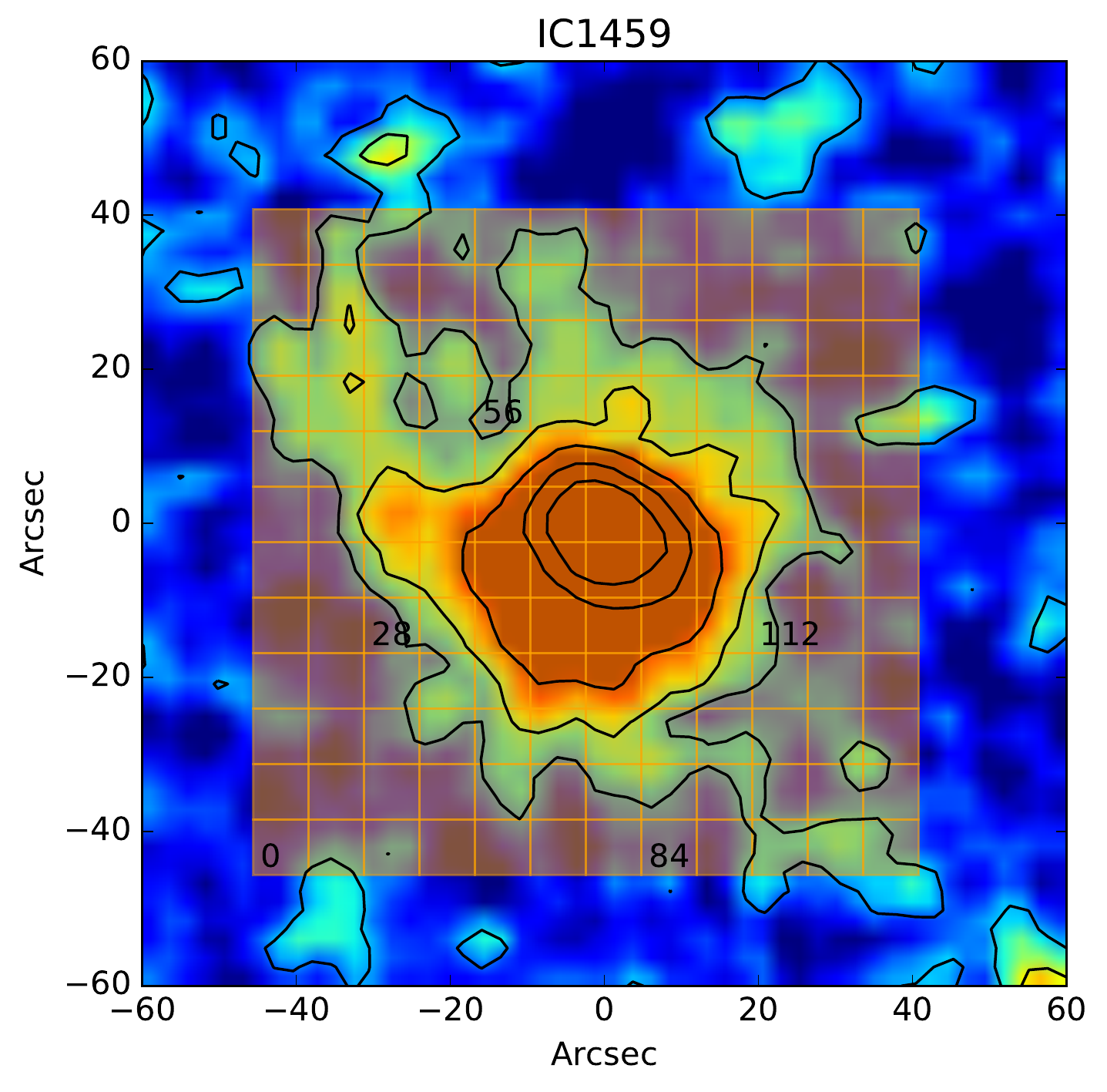}
\caption{{\bf Left :} 2MASS K-band image of IC1459 with the grid of our 7.2'' pixels superimposed on top. {\bf Right :} PACS 160$\mu$m image of IC1459 with the grid of our 7.2'' pixels superimposed on top. The black numbers in the grid represent the pixel index, it increases from bottom to top and left to right (in the RADEC coordinate system, the pixel index goes from south to north then east to west).}
\label{fig:icgrid}
\end{center}
\end{figure*}

\subsection{NGC2768 analysis}

\begin{figure*}[h!t]
\begin{center}
\includegraphics[width=8cm]{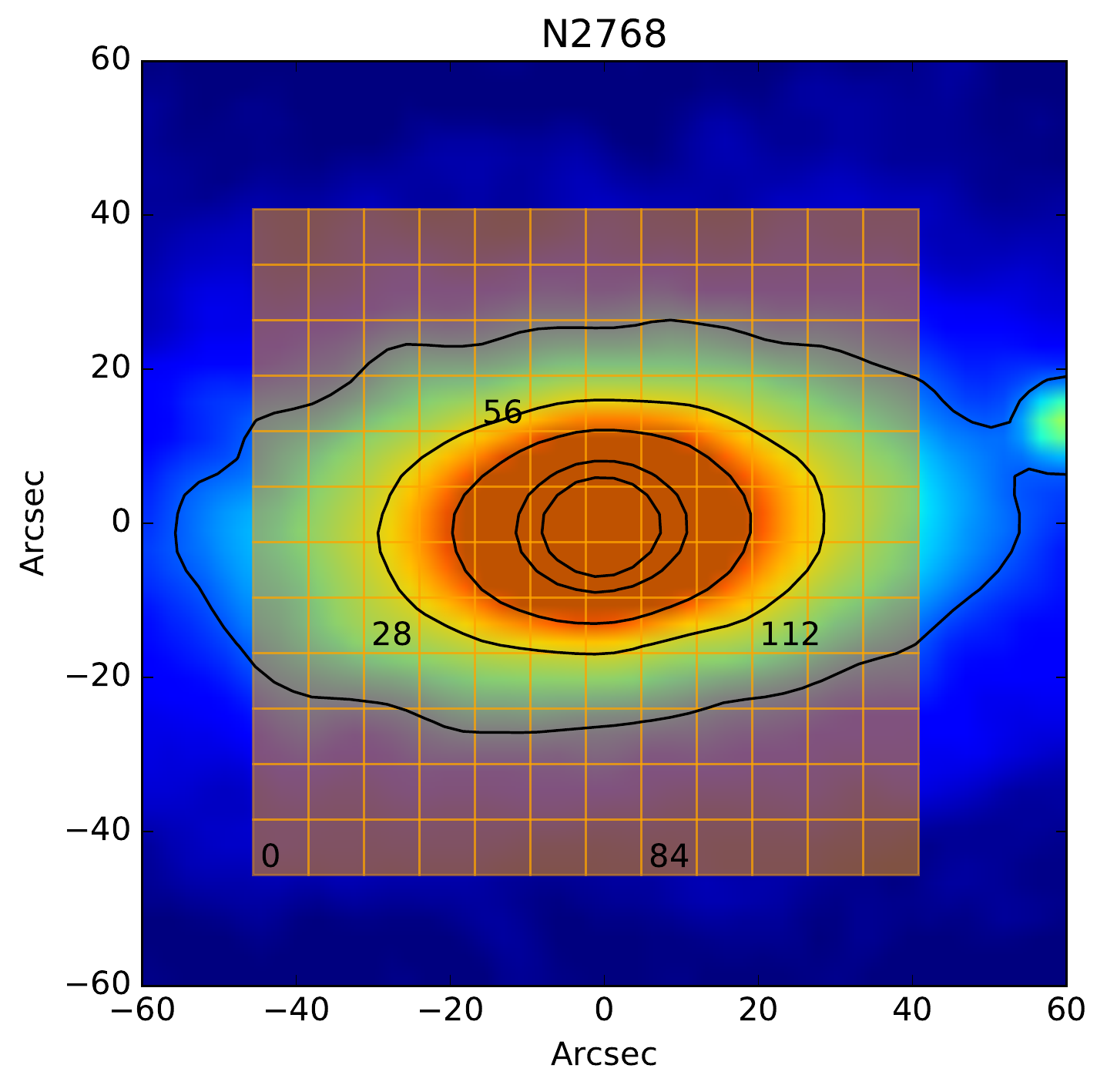}\hspace{0.5cm}\includegraphics[width=8cm]{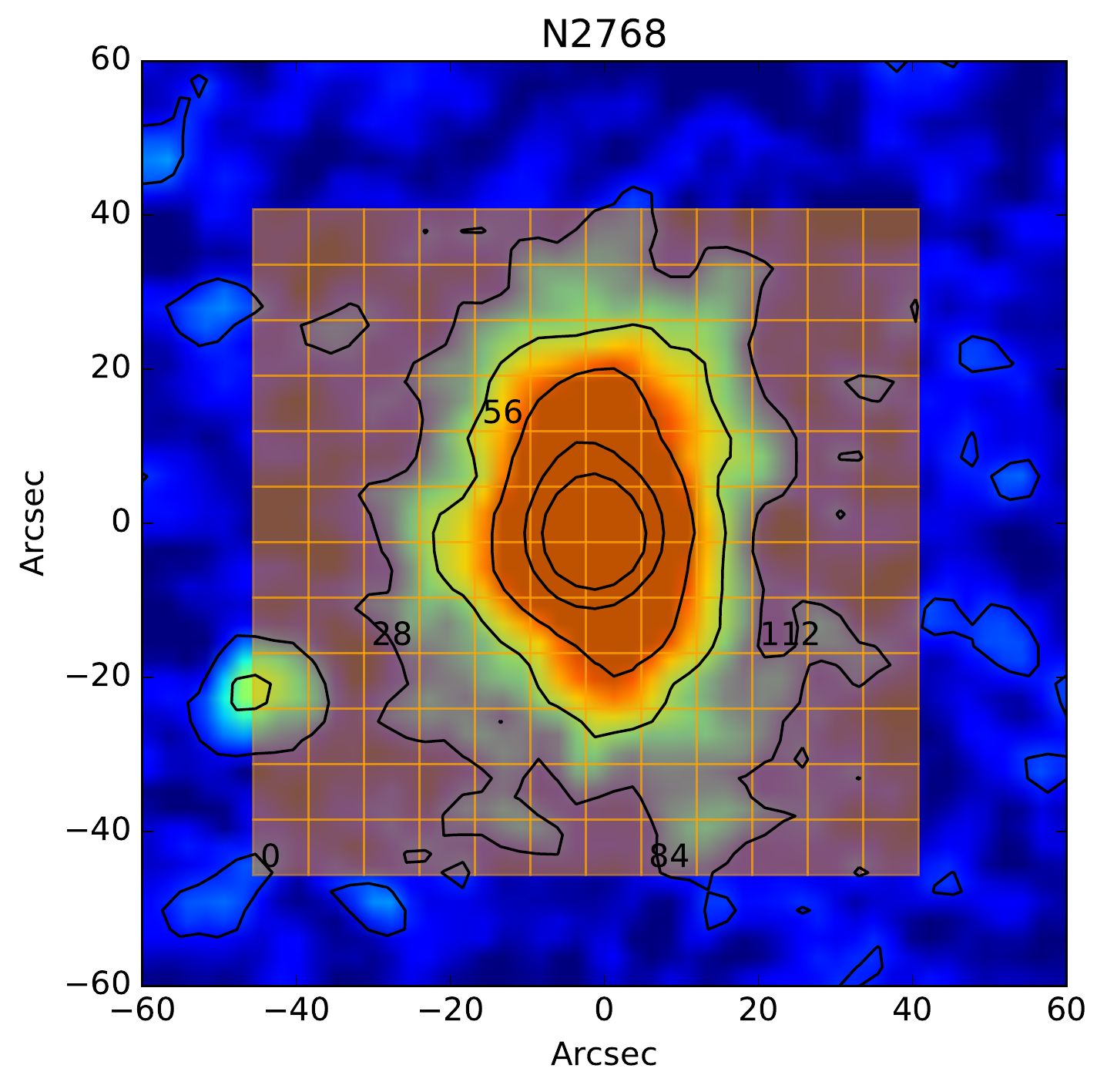}
\caption{{\bf Left :} 2MASS K-band image of NGC2768 with the grid of our 7.2'' pixels superimposed on top. {\bf Right :}  PACS 160$\mu$m image of NGC2768 with the grid of our 7.2'' pixels superimposed on top. The black numbers in the grid represent the pixel index, it increases from bottom to top and left to right (in the RADEC coordinate system, the pixel index goes from south to north then east to west).}
\label{fig:ngrid}
\end{center}
\end{figure*}

At optical-NIR wavelengths the NGC2768 morphology resembles an ellipsoid elongated in the RA direction,
whereas in the MID to FIR the elongation migrates towards the DEC direction.
The reported difference in the elongation direction is in agreement with observation of a polar (along minor axis)
ring/dust lane observed by \cite{kim89} in B-R color and by \cite{crocker08} in CO 
line emission. The CO lines show as well
that this ring/disk is rotating (toward us in the south and away from us in the north). 
These results match observation of the [OIII] line from \cite{sarzi06} showing the same
 rotation of the ionized gas.\\
To understand the difference in gas and star content within NGC2768, we split the 
galaxy into 144 area, each covering 7.2$\times$7.2 arcsecond$^2$. In figure \ref{fig:ngrid}, the 
grid of our 144  pixels are superimposed on a map of NGC2768 from the 2MASS K-band
and the PACS 160$\mu$m band. The black numbers on the maps represent the pixel index, that
increases from bottom to top and left to right (in the RADEC coordinate system, south to north
then east to west).

\begin{figure*}[h!t]
\begin{center}
\includegraphics[width=16cm]{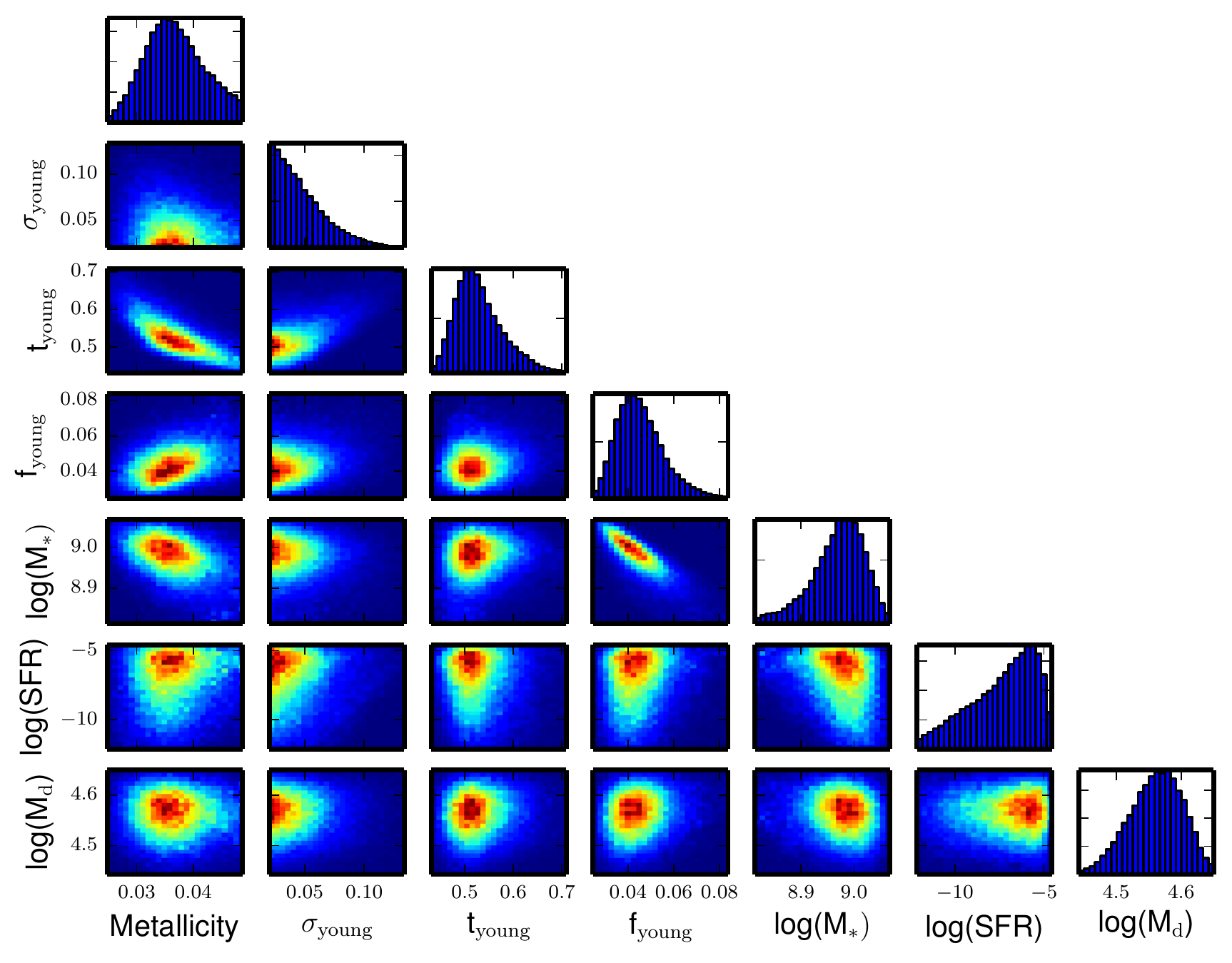}\\
\caption{Constraints on the metallicity, $\sigma_{young}$, t$_{young}$, f$_{young}$, log(M$_*$), log(SFR) and log(M$_d$)
of the pixel 80 of NGC2768. Most parameters plotted in this figure do not correlate strongly, the parameter pairs
(t$_\mathrm{young}$, Metallicity) and (f$_\mathrm{young}$,log(M$_*$)) are slightly anti-correlated.}
\label{fig:tri}
\end{center}
\end{figure*}

\begin{figure*}[h!t]
\begin{center}
\includegraphics[width=18cm,height=6cm]{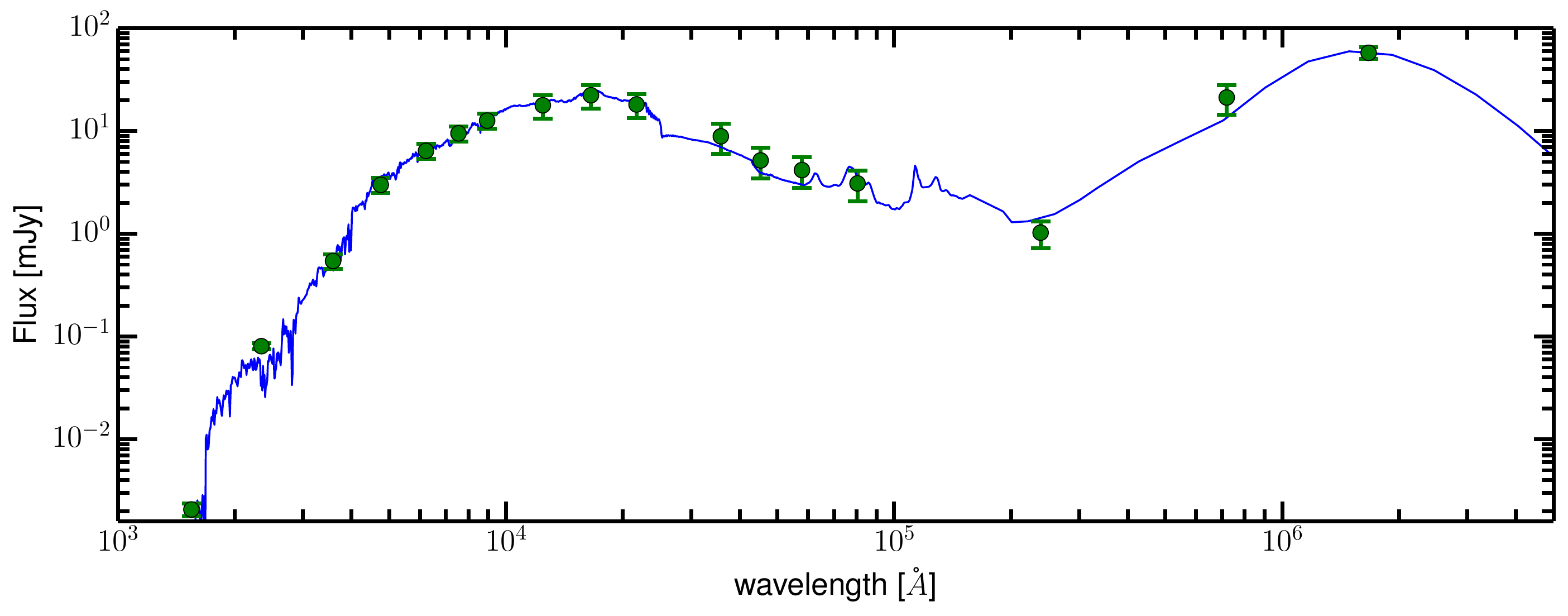}\\
\caption{SED of the pixel 80 of NGC2768, the blue solid line represents the best modeled SED, green dots represent data observed with GALEX, VST ATLAS, 2MASS, {\it Spitzer} and {\it Herschel}.}
\label{fig:sed}
\end{center}
\end{figure*}

Each pixel is then fitted with our SED model, and model parameters can be mapped throughout
the galaxy. The constraints shown on each plot have been marginalized over the remaining
set of parameters. Some parameters are poorly constrained such as the age of the old stellar 
population and the old population e-folding time. Some others only have upper or lower
limits such as the far-IR luminosity fraction contributed by an AGN, or the duration of the recent star burst. 
Lastly some parameters, like $\alpha$ and the age of the new stellar population, 
while well constrained, do not present a statistically significant
variation across the map, and only their average value will be reported.\\ 
Figure \ref{fig:tri} represents a set of constraints on a pixel of NGC2768 (pixel \#80, slightly 
above the central pixel, [x,y] coordinates [6,8]) and show
the correlation between parameters. Overall most parameters are not very correlated. The 
strongest correlation is  the anti-correlation between the 
fraction of young stars and the stellar mass. The other one is the anti-correlation between the
metallicity and the age of the young stellar population. These correlations widen our uncertainty
on these parameters as constraints presented in this paper are marginalized over the value of all the 
other parameters.\\
The metallicity is one of the major constraint coming from our fluxes and models, in figure \ref{fig:metn}
it is shown to decrease sharply with the radial distance from the center of the galaxy. The uncertainty in
our metallicity measurements is from 0.003 to 0.005 (7 to 16\%). The shape of 
its distribution follow the old stellar population, which is dominant in NGC2768. The overall values of the 
metallicity, between 0.025 and 0.045, is in good agreement with measurements from \cite{silchenko06,howell05,denicolo05}, ranging from 0.02 to 0.048. Their measurements used Lick indices and SSP models from \cite{thomas03}
to extract the age and metallicity of NGC2768 within a radius of about 8''.
\cite{li07} measured a metallicity of 0.031$\pm$0.009 using B-K and B-V colors within 
the effective radius and \cite{bruzual03} SPS models with a Salpeter IMF.
These results are also qualitatively in agreement with \cite{foster09}, where the Ca triplet 
lines (between 8483 and 8677 Angstrom) were used to probe the metallicity from \re/8 to 2\re (\re $\simeq$ 64'') with the models of \cite{vazdekis03} using a Kroupa IMF. 
\cite{foster09} measured a lower metallicity, ranging from 0.002 to 0.02, but also found a decreasing
metallicity with radius. 
Their slope in the first \re$\,$  is about -0.015 whereas ours is about -0.019 in Z
(in $\delta$log(Z)/$\delta$log(r/r$_e$) our slope is about -0.21, whereas theirs is about -0.6). 
\cite{kuntschner10}, using Lick indexes and SSP models from \cite{thomas03}, found a decreasing 
metallicity gradient going from 0.02 at \re/8 to 0.013 at \re, corresponding to a -0.007 gradient in Z.

\begin{figure*}[h!t]
\begin{center}
\includegraphics[width=8cm]{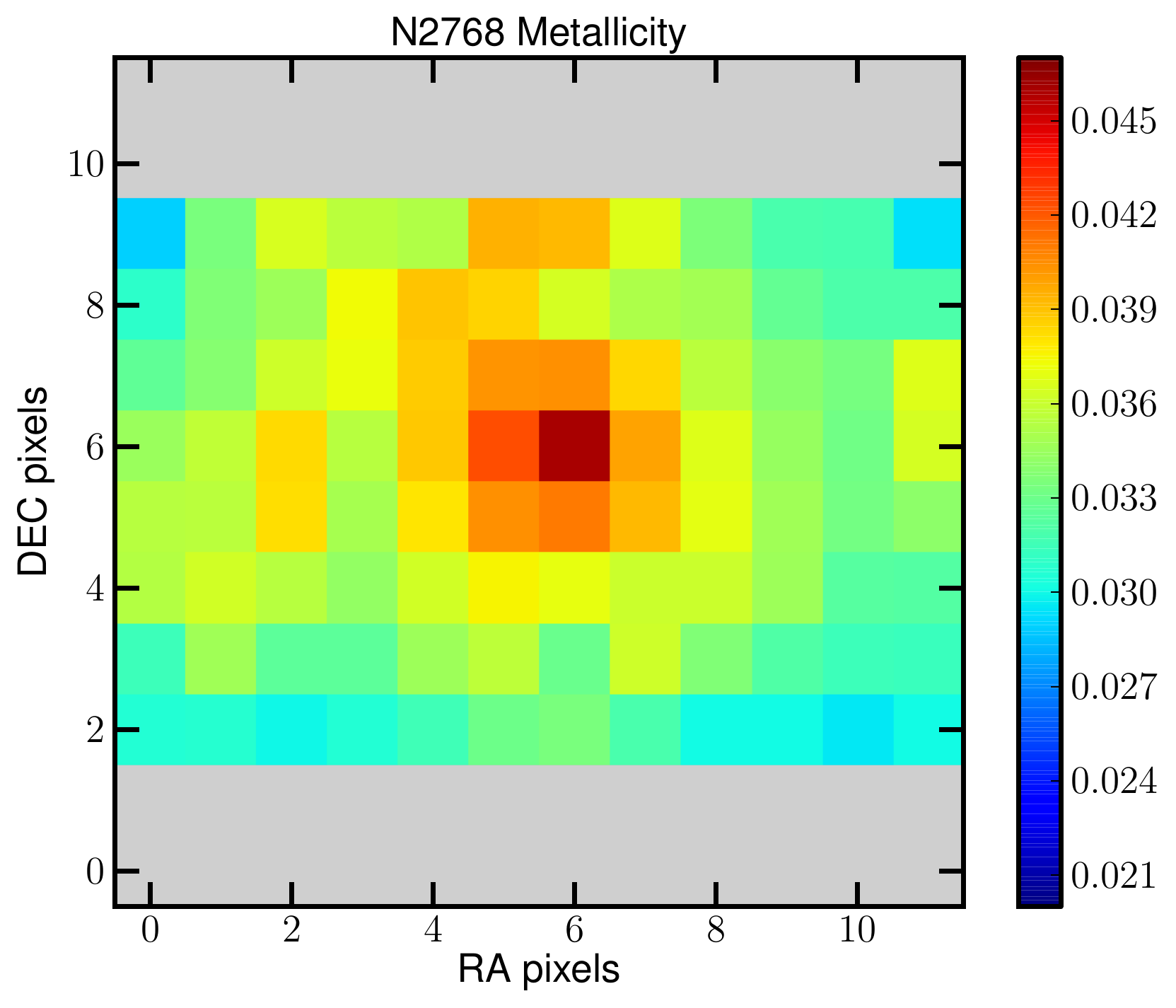}\hspace{0.5cm}\includegraphics[width=8.6cm]{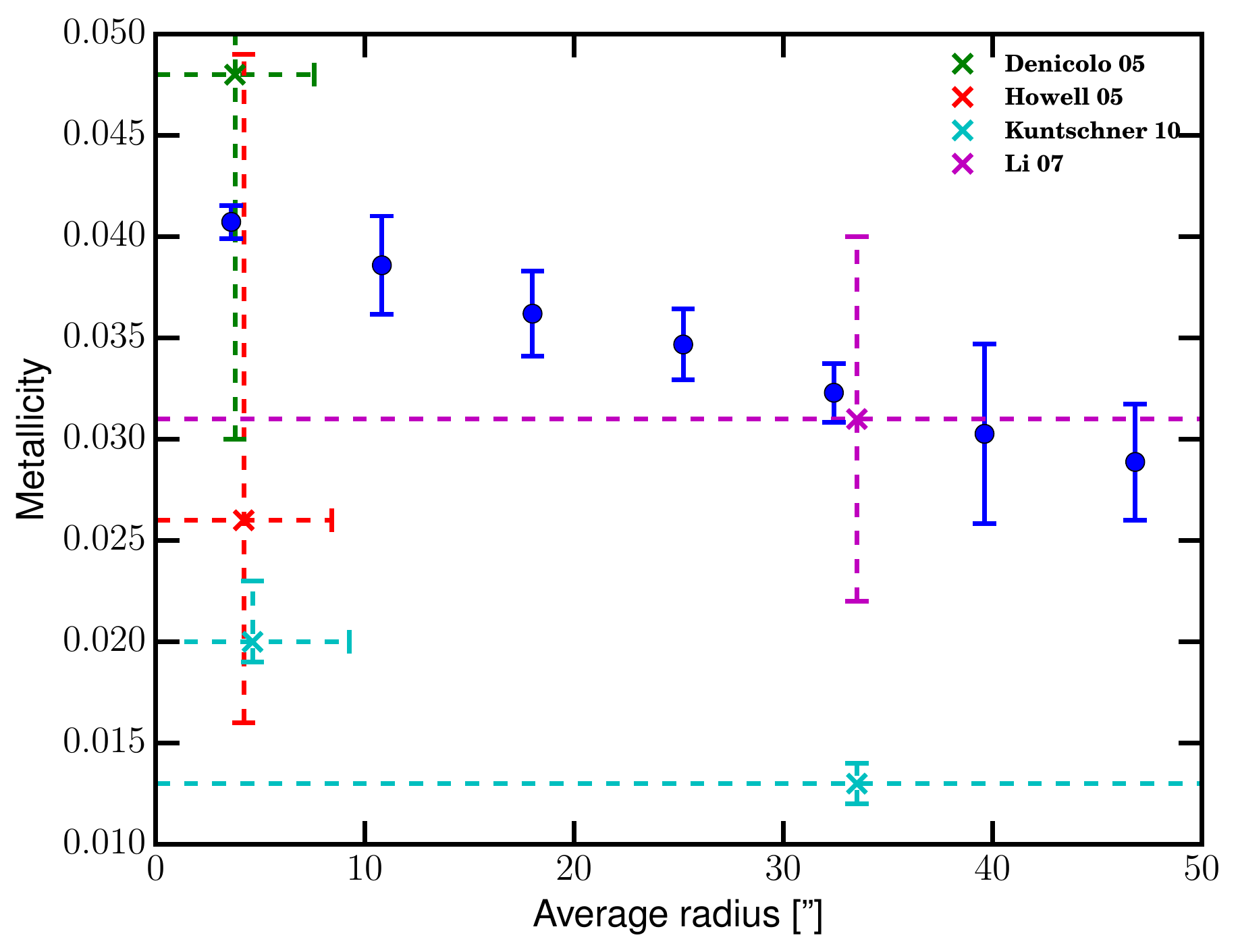}\\
\caption{{\bf Left :} Metallicity of NGC2768 within the 7.2'' pixel grid, unconstrained pixels are colored in gray, the average error on the value of each pixel is about 0.004. {\bf Right :}  radial profile of the metallicity of NGC2768 (\re = 64'') with previous measurements from \cite{denicolo05,howell05,kuntschner10,li07}. Error bars of our radial profile indicate
the spread of the metallicity within the radial bin.}
\label{fig:metn}
\end{center}
\end{figure*}

The fraction of young star fitted across NGC2768 is on average about 3.5\% and peaks at about 5\% in the polar regions
where the dust ring is located. The central pixel of NGC2768 has a young star fraction of about 2\%.
The average error on the determination of the young star fraction is fairly substantial at about 1\%. 
The average age of the young stellar population is about 0.5$\pm$0.05 Gyr, the duration (1 $\sigma$ of the Gaussian-shaped burst) 
of the recent burst of star formation is poorly constrained and only an average upper limit of about on 100 Myr 
(95\% confidence level) is derived. \cite{kuntschner10} did not detect a younger stellar population
using H$\beta$ absorption, but the low sensitivity of H$\beta$ absorption to a small fraction and recent
star formation \citep{crocker08} could explain the difference with our results. The polar ring region shows a slightly
larger age for the young stellar population (0.55 Gyr) but it is not significant given our uncertainties.\\

The age and duration of the old stellar population formation is unconstrained and marginalized over the prior
range, from 8 to 14 Gyr for its age and 0.5 to 3 Gyr for its e-folding time. Other studies have calculated an average
age of the stellar population of NGC2768, the results vary widely from 4 Gyr to 11 Gyr. \cite{silchenko06} derived  a luminosity-weighted mean age of 11 Gyr for the galaxy nucleus, \cite{serra08} found an age of 4 Gyr,
\cite{zhang08} estimated an age of about 6 Gyr. \cite{howell05} derived an age of 10 Gyr with a large 7 Gyr uncertainty and
\cite{denicolo05} calculated an age of 8.0 $\pm$ 3.5 Gyr, both within re/8.

\begin{figure*}[h!t]
\begin{center}
\includegraphics[width=8cm]{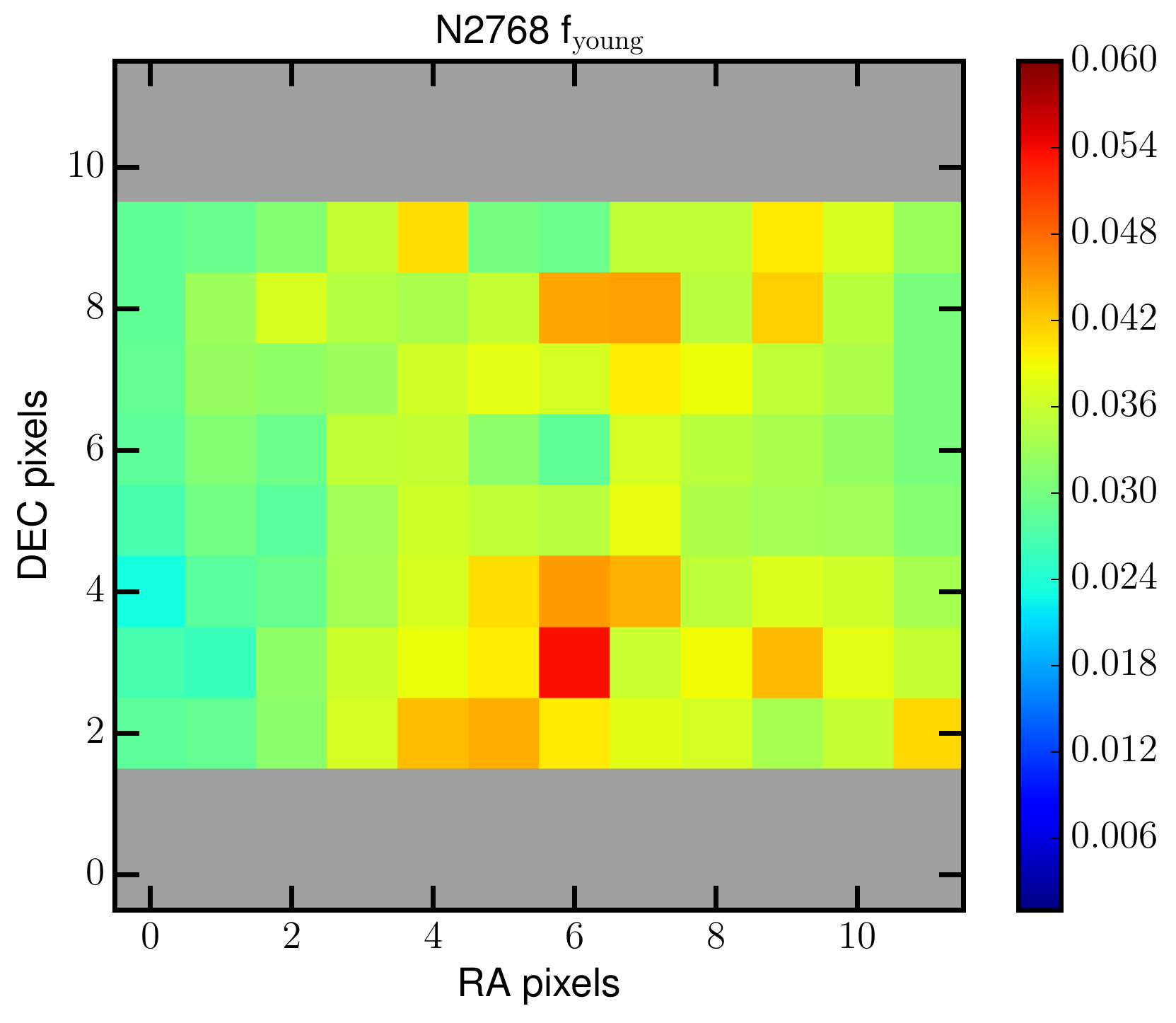}\hspace{0.5cm}\includegraphics[width=8.6cm]{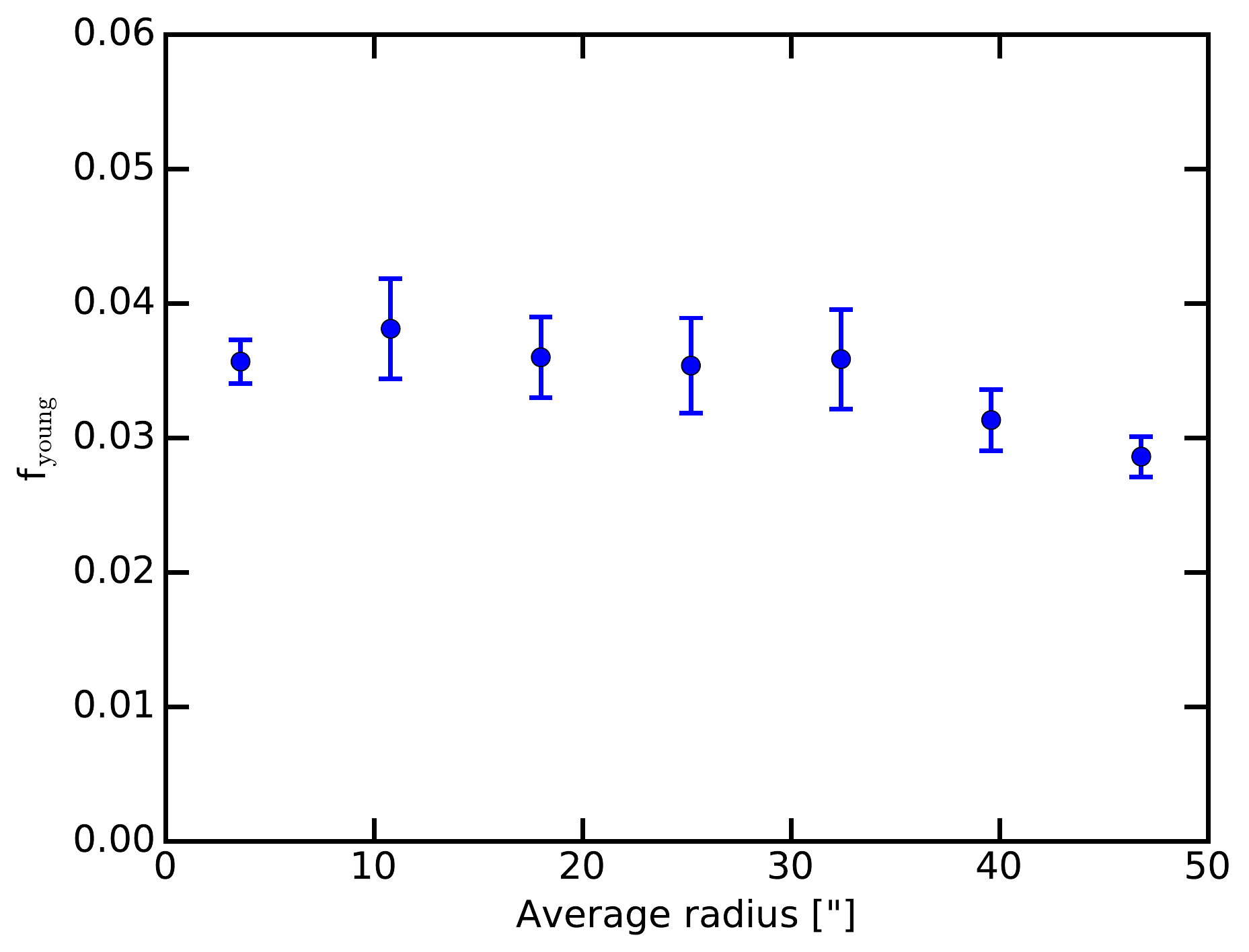}\\
\caption{{\bf Left :} young star fraction of NGC2768 within the 7.2'' pixel grid, unconstrained pixels are colored in gray, the average error on the value of each pixel is about 0.01. {\bf Right :}  radial profile of the young star fraction of NGC2768 (r$_e$ = 64''), error bars indicate the spread of the fraction within the radial bin.}
\label{fig:fn}
\end{center}
\end{figure*}

\begin{figure*}[h!t]
\begin{center}
\includegraphics[width=8cm]{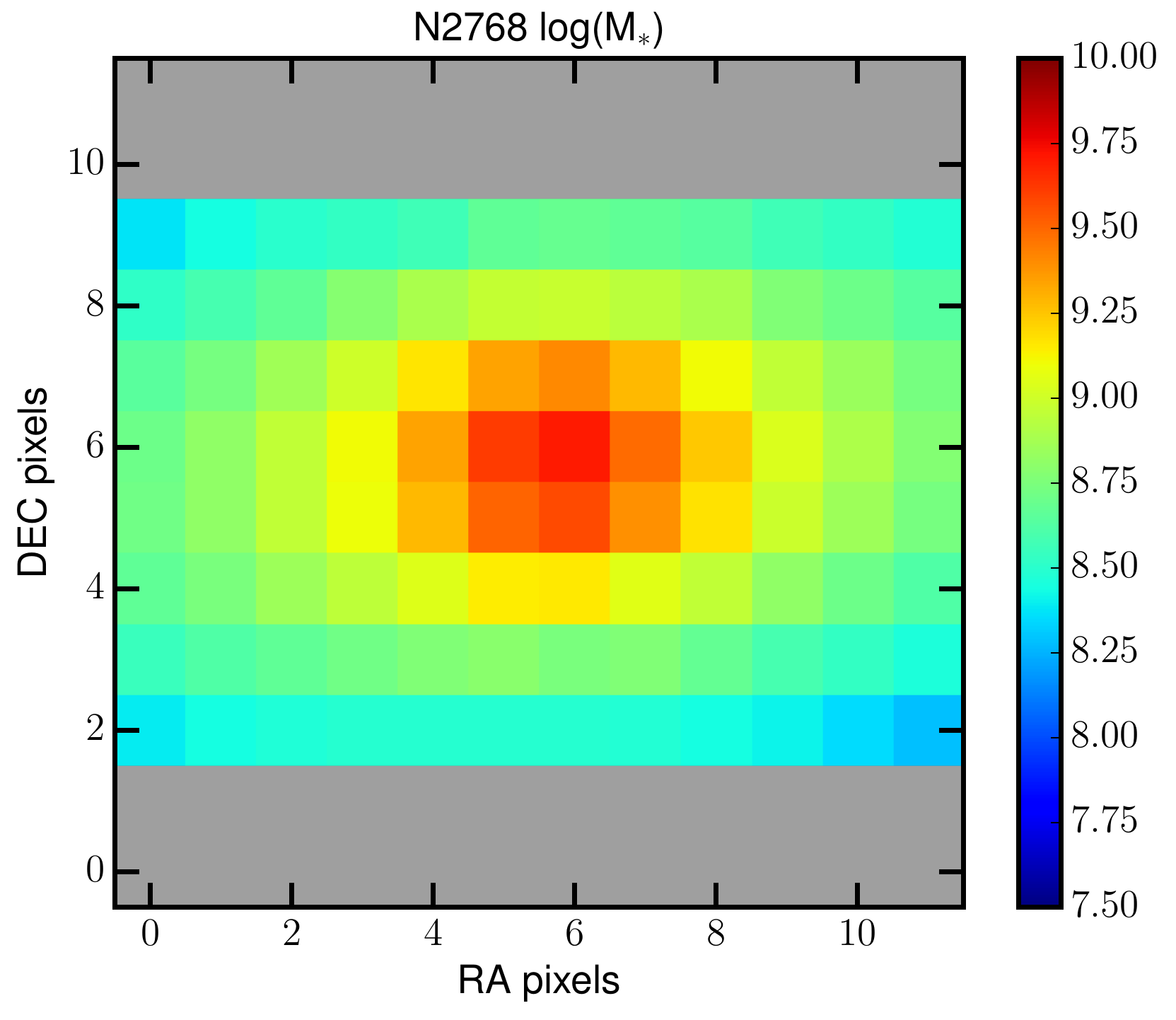}\hspace{0.5cm}\includegraphics[width=8.2cm]{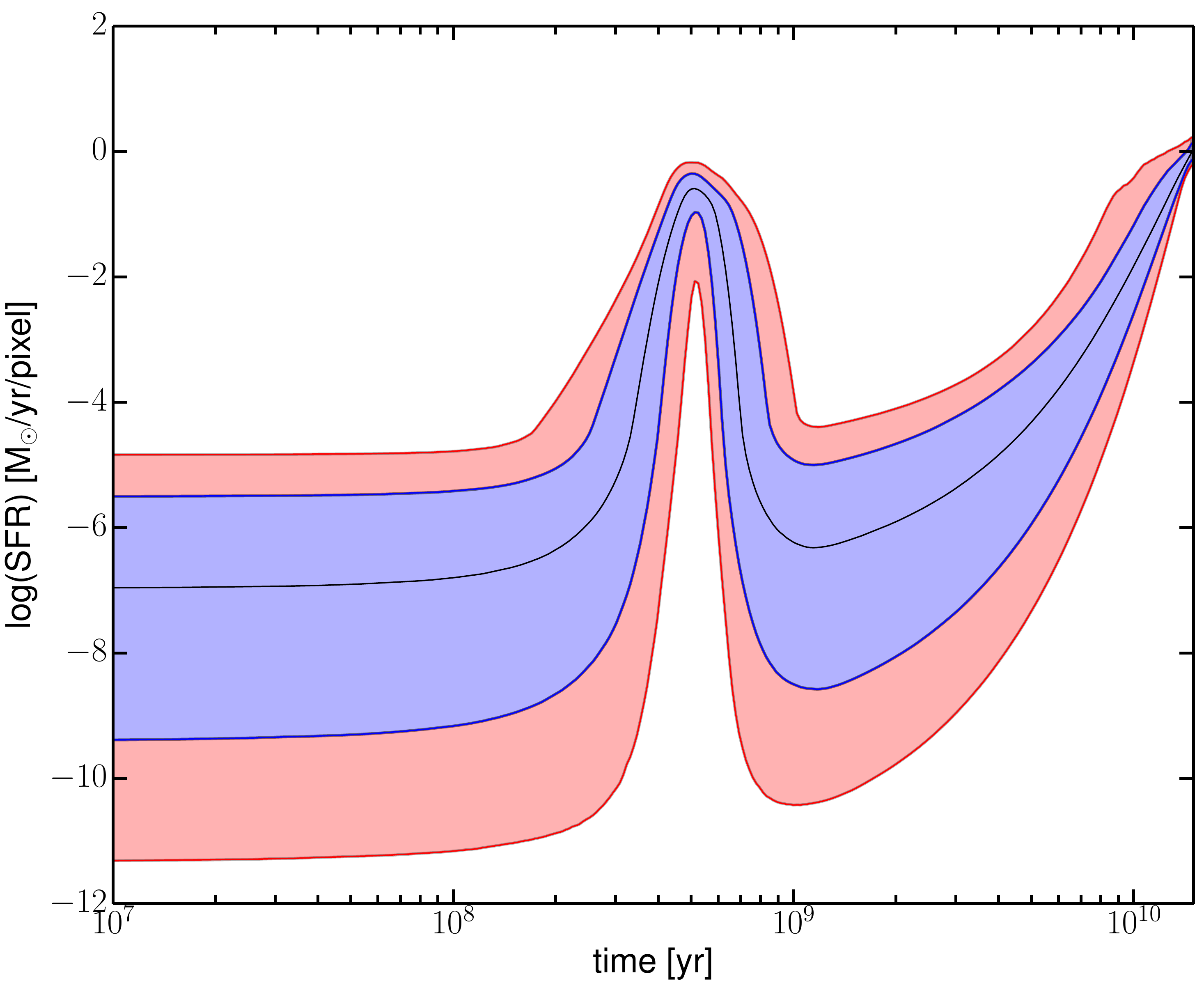}\\
\caption{
{\bf Left :}  logarithm of the stellar mass of NGC2768 within the 7.2'' pixel grid, unconstrained pixels are colored in gray, the average error on the value of each pixel is about 0.04. {\bf Right :} Star formation history of pixel \#80 as modeled with a Gaussian-shaped burst and an exponential decrease, the blue contour represent the 68\% confidence interval and the red contour represents the 95\% confidence interval. The star formation rate is best constrained around the peak of the burst.}
\label{fig:sfrn}
\end{center}
\end{figure*}

f$_{AGN}$ which represents the fractional contribution of a model of AGN to the total FIR luminosity, is constrained to be lower than 2\% across the entire galaxy at a 95\% confidence level. 
\cite{dale02} model 
parameter $\alpha$ can only be given a lower limit of about 2.7 across the galaxy as well at a 95\% confidence level.
The recent SFR  is not very well constrained because the dominant term in the SF history model 
is the old stellar population term which parameters are not well constrained. Given the amplitude of the 
uncertainties the residual SFR is the same across the galaxies at a rate of 10$^{-7.5\pm2.0}$ M$_\odot$/yr/pixel.\\ 
Figure \ref{fig:sfrn} shows the reconstructed star formation history at the location of  pixel \# 80 in our parameter
maps, the blue and red contours indicate respectively the 68\% and 95\% confidence levels. The t=0 SFR is not
well constrained and is dominated by the residual star formation from the exponential decrease. This SFR component is not well correlated with the dust distribution, which follows the SFR of the burst of young stars.
The burst of young stars at about 0.5 Gyr is detected at a significant level above the expected old stellar population SFR. The amplitude of this burst is better constrained than the present day SFR.\\
The dust absorption A$_\mathrm{V}$ is constrained at the center of the galaxy along the polar ring
as seen in figure \ref{fig:avn}. A$_\mathrm{V}$ peaks at about 0.07 mag in the galaxy center and decreases
rapidly away from the polar ring. Further away the 95\% confidence level upper limit is about 0.013.
The dust absorption in V band is well correlated with the PACS 160$\mu$m image indicating
a good consistency between the dust absorption near optical wavelengths and emission in the far-infrared.
Figure \ref{fig:mdn} reveals a dust mass distribution that is, as expected, also correlated with the PACS 160$\mu$m image. The dust mass radial profile slope is about -0.09 dex/arcsecond.

\begin{figure*}[h!t]
\begin{center}
\includegraphics[width=8cm]{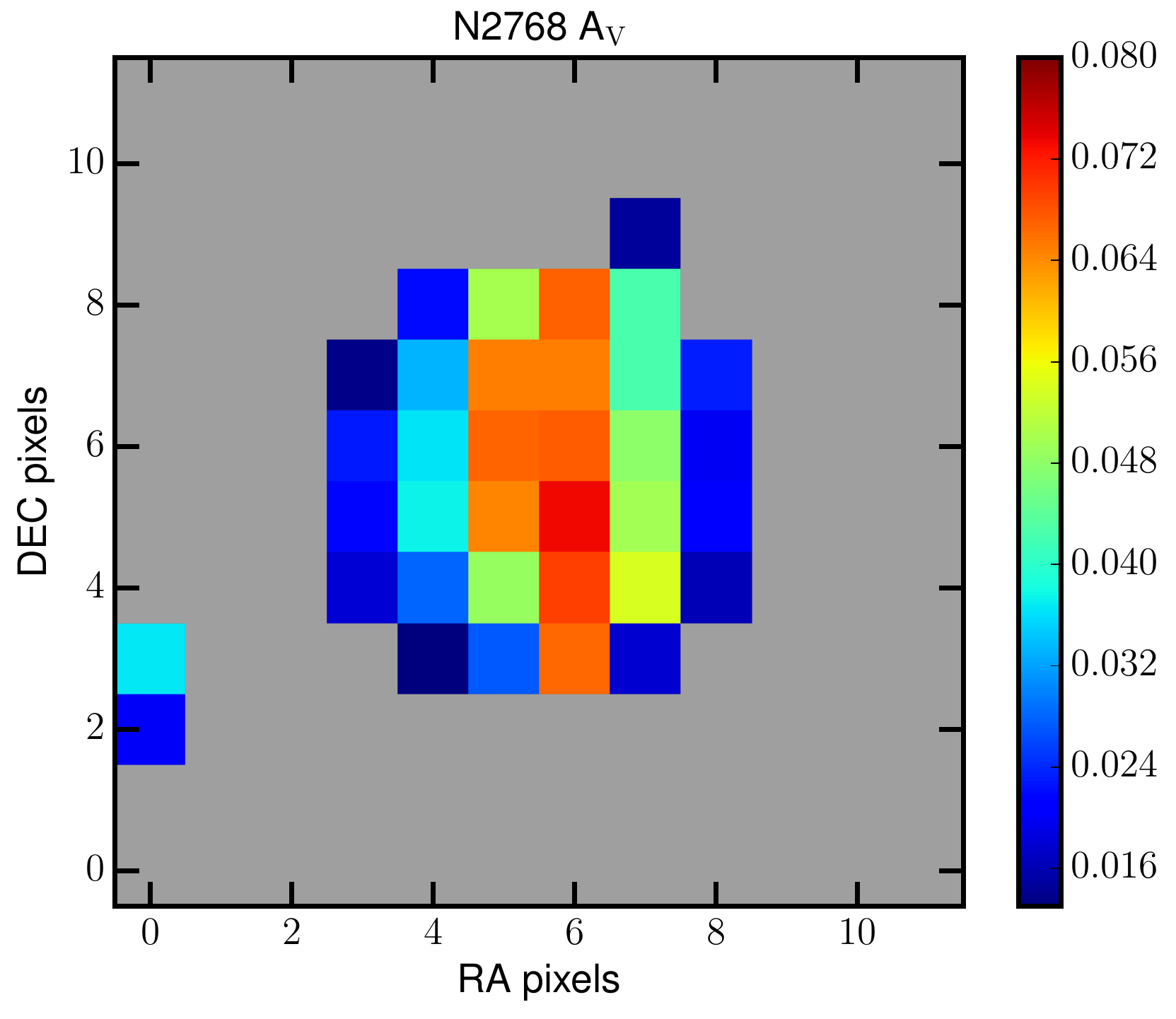}\hspace{0.5cm}\includegraphics[width=8.6cm]{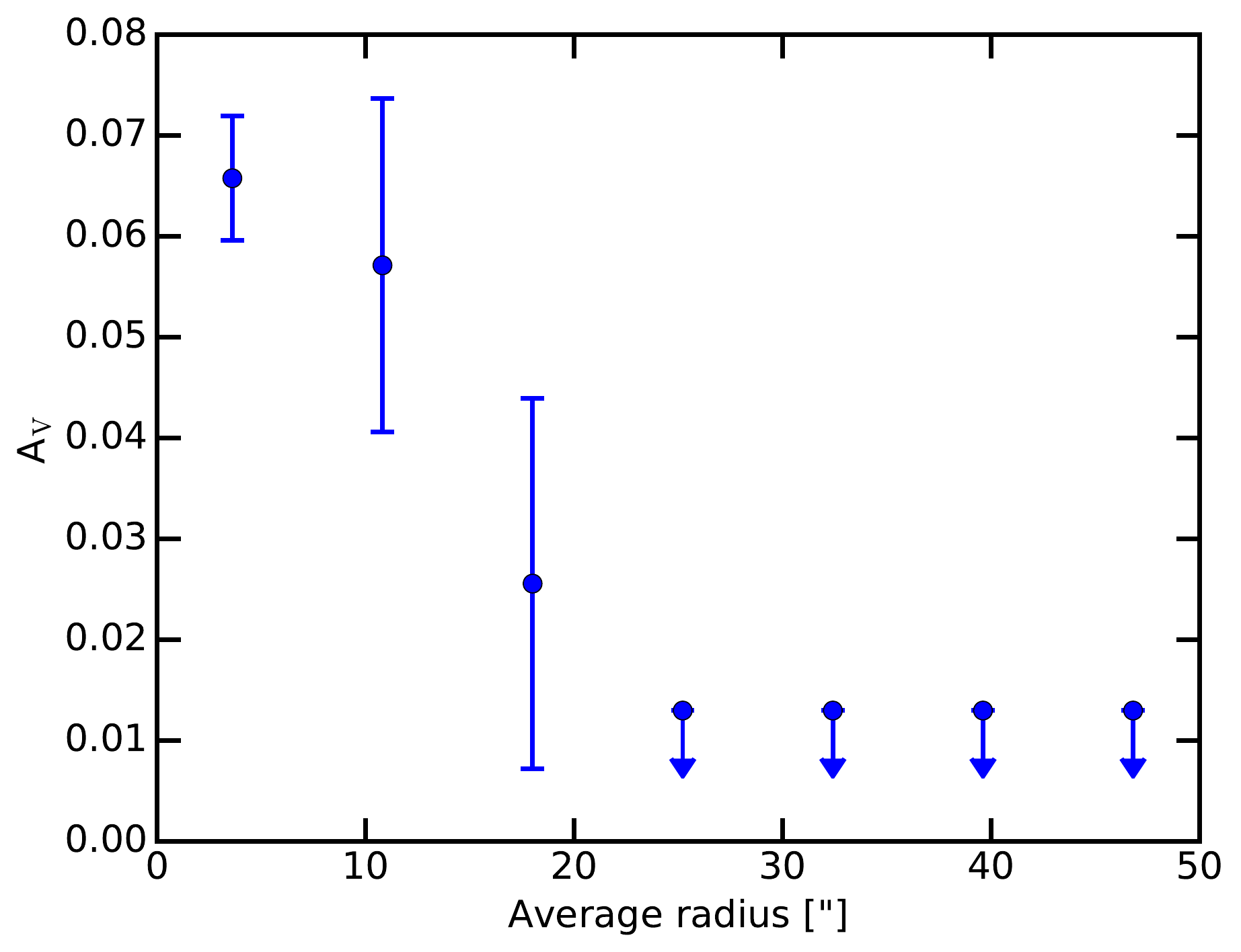}\\
\caption{{\bf Left :} dust absorption in the V band of NGC2768 within the 7.2'' pixel grid, unconstrained pixels are colored in gray, the average error on the value of each pixel is about 0.005. {\bf Right :}  radial profile of dust absorption of NGC2768 (r$_e$ = 64''), arrows indicate upper limits and error bars indicate the spread of the dust absorption within each radial bin.}
\label{fig:avn}
\end{center}
\end{figure*}
\begin{figure*}[h!t]
\begin{center}
\includegraphics[width=8cm]{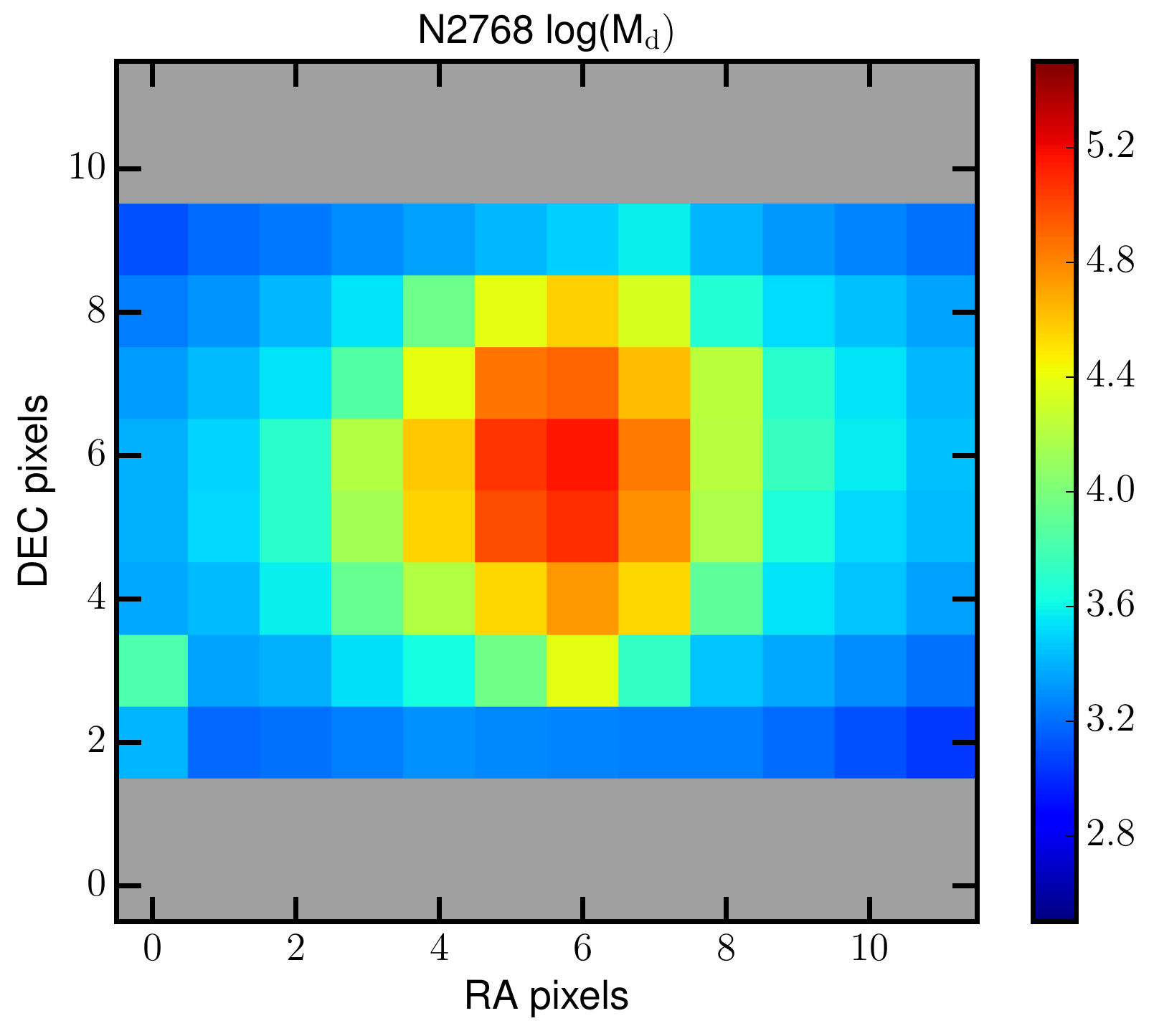}\hspace{0.5cm}\includegraphics[width=8.6cm]{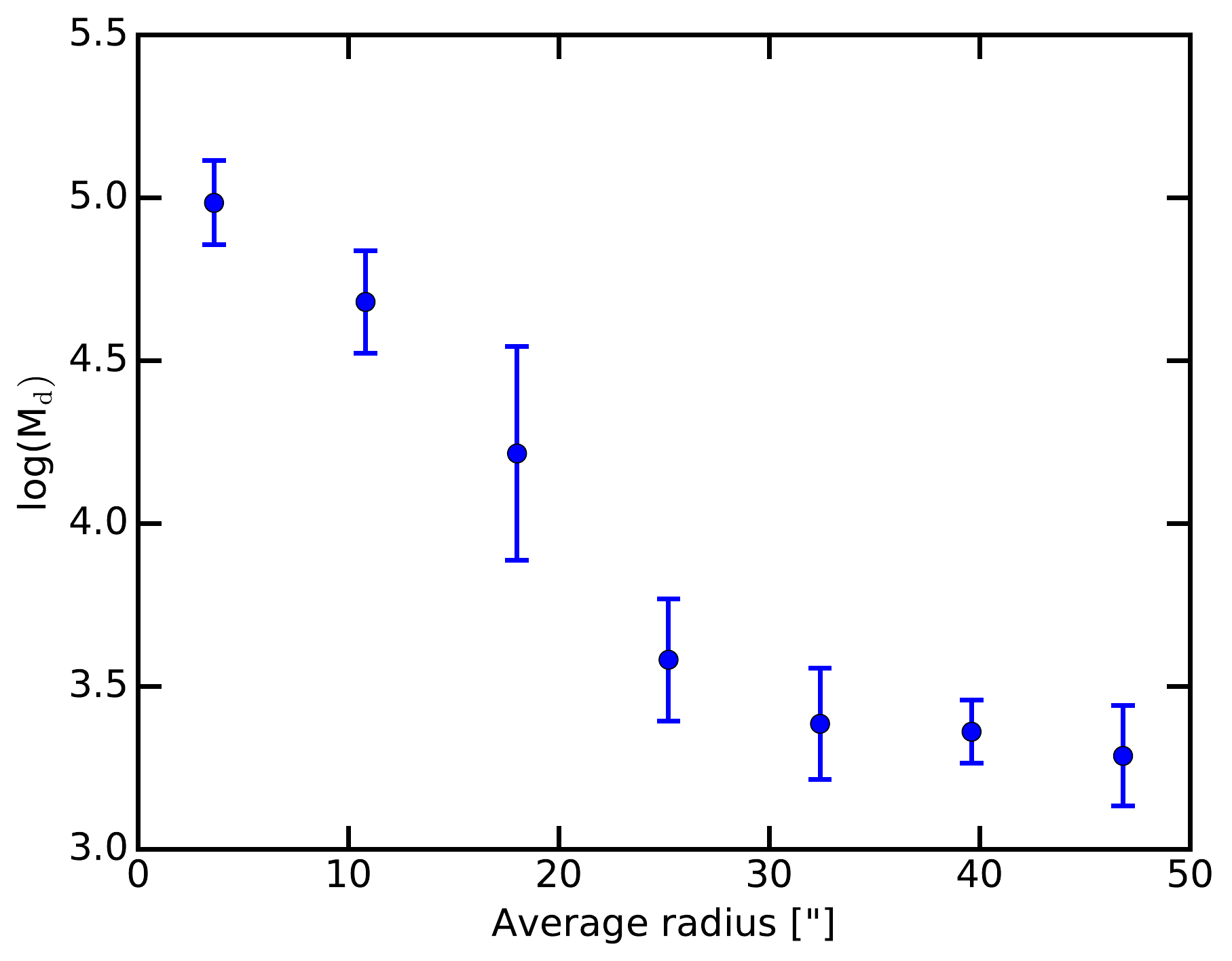}\\
\caption{{\bf Left :} dust mass  of NGC2768 within the 7.2'' pixel grid, unconstrained pixels are colored in gray, the average error on the value of each pixel is about 0.06. {\bf Right :} dust mass of NGC2768 (r$_e$ = 64''), error bars indicate the spread of the dust mass within each radial bin.}
\label{fig:mdn}
\end{center}
\end{figure*}

\subsection{IC1459 analysis}

IC1459 has a very regular ellipsoidal morphology elongated along the north-west direction in the K band 
(fig. \ref{fig:icgrid}), but its PACS 160 $\mu$m image shows that 
the gas does not follow the stellar content and has a much more disturbed morphology.
In addition to the dust located at its center, IC1459 has large strip of dust lining from the south-east
to the north-west on its west side. \cite{Goudfrooij90} identified some potential spiral arms in H$\alpha$
maps of IC1459 that extend about 40'' away from its center. 
The location of the two largest H$\alpha$ arms seems to coincide with the location of the structures 
present north of the galaxy center in the 160$\mu$m map, although the FIR structures do not resemble closely the shape of spiral arms.\\
To differentiate the evolution and content of different regions of IC1459, we split the 
galaxy into 144 area, each covering 7.2 arcsecond$^2$ in the same fashion as
for NGC2768. In figure \ref{fig:icgrid}, the grid of our 144  pixels are superimposed on a map of 
IC1459 from the 2MASS K-band and the PACS 160$\mu$m band. The black numbers on the maps represent the pixel index, that
increases from bottom to top and left to right (in the RADEC coordinate system, from south to north then east to west). Each pixel is then fitted with our SED model like for 
NGC2768 and constraints shown on each plot have been marginalized over the remaining set of parameters.\\

The fitted metallicity map (fig. \ref{fig:meti}) reveals high metallicity values roughly along the direction of 
the major axis of IC1459, where metallicity is on average 0.044 compared to a metallicity of
about 0.042 for neighboring pixels. On the other hand, the south west side of this strip has a 
lower metallicity than its opposite side (0.037 vs 0.042). This south west side corresponds
to the location of the extended structure in PACS 160$\,\mu$m map.
Overall IC1459 metallicity is decreasing with the radial distance from its center beyond 15'' at a rate of -0.012 per effective radius in Z (about -0.36 in [Z/H]
and about -0.39 in the $\delta$log(Z)/$\delta$log(r/r$_e$)).
\cite{annibali07}, using Lick indices and a SSP model,
 found a metallicity of 0.042 $\pm$ 0.009 at the center of IC1459 (r < re/8) 
. \cite{cappellari02} found a color gradient that could be explained by a metallicity gradient.
\cite{li07} measured a metallicity of 0.039$\pm$0.004 using B-K and B-V colors within 
the effective radius and \cite{bruzual03} SPS models with a Salpeter IMF. Our measured metallicity
profile  is shown in figure \ref{fig:meti} to be in good agreement with these previous measurements.

\begin{figure*}[h!t]
\begin{center}
\includegraphics[width=8cm]{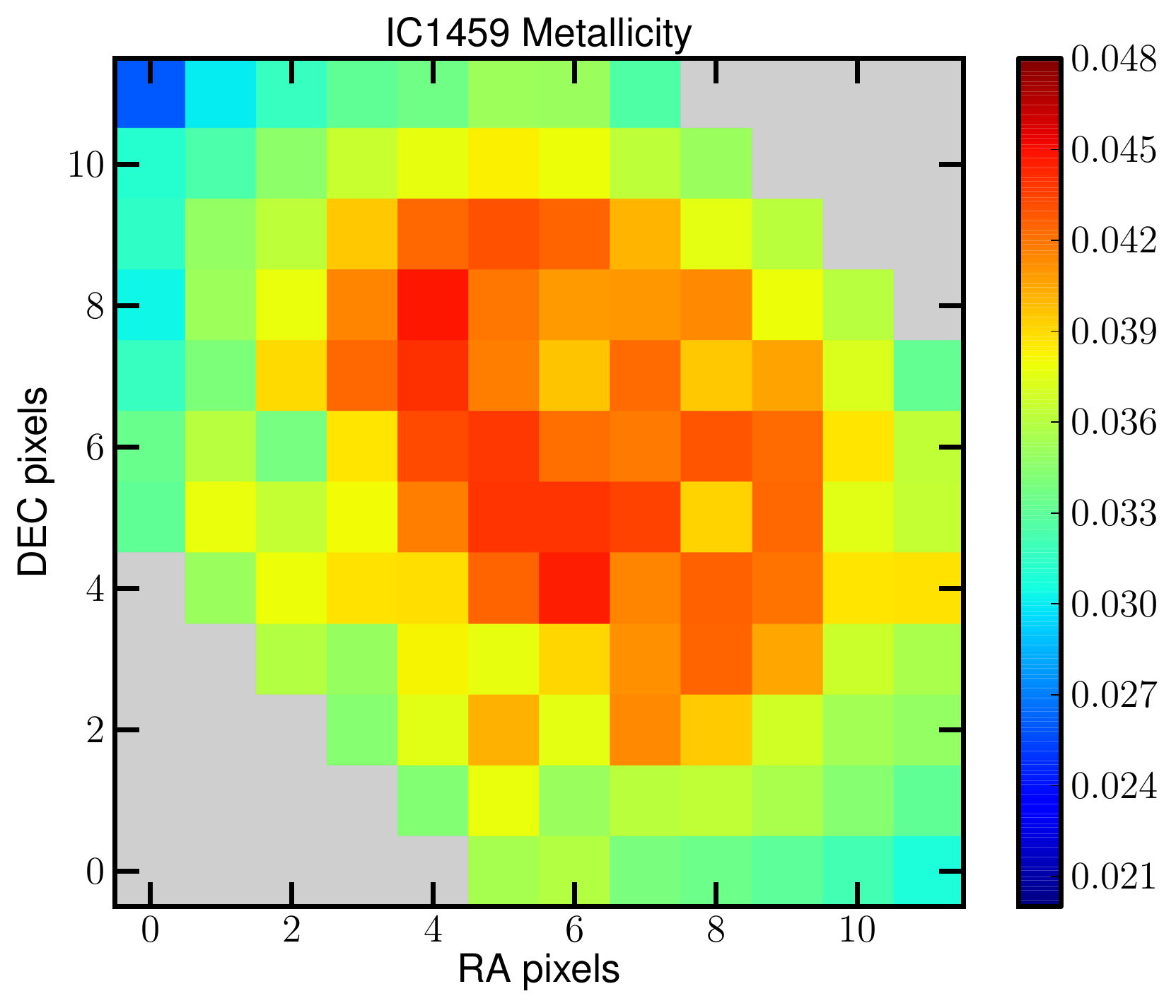}\hspace{0.5cm}\includegraphics[width=8.6cm]{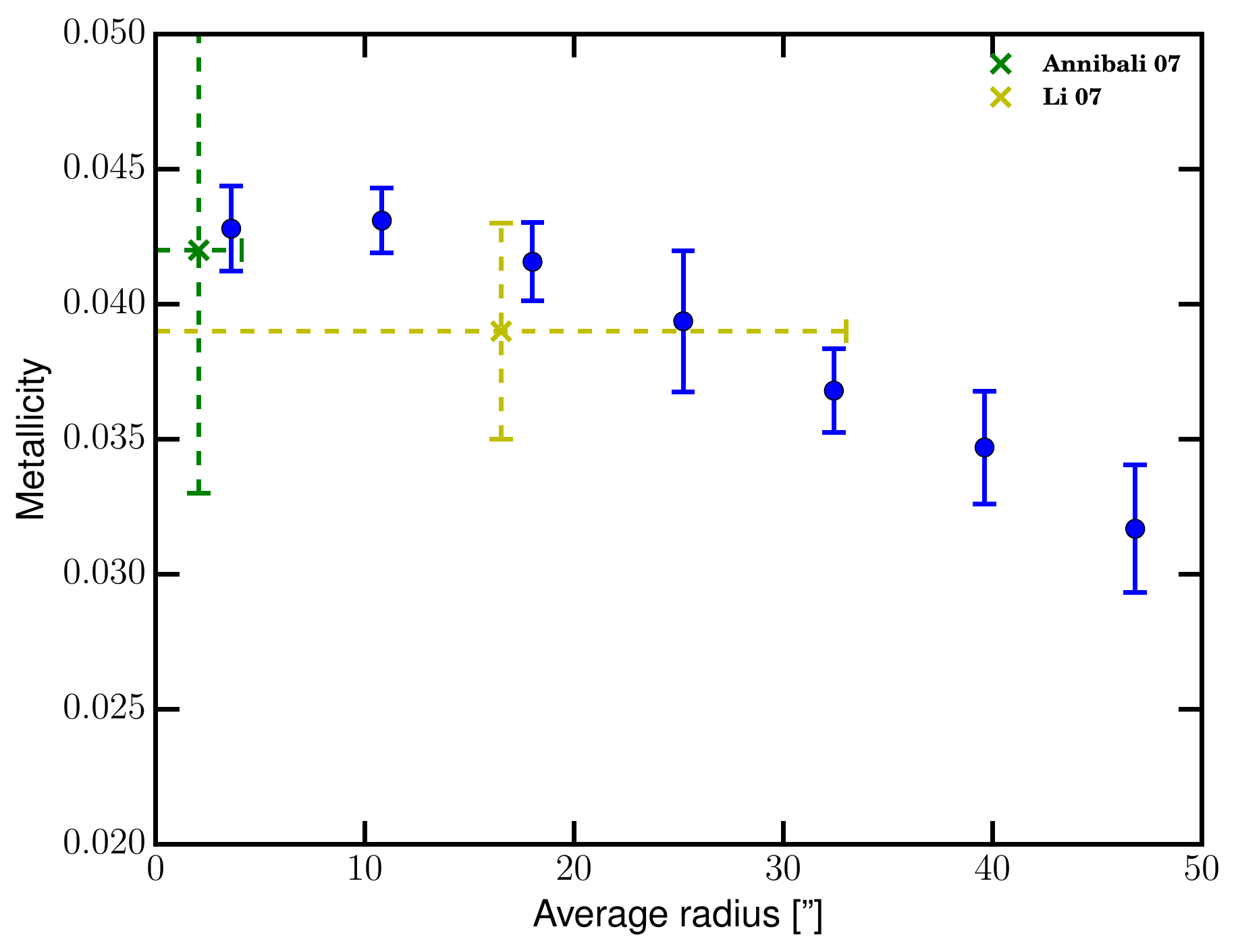}\\
\caption{{\bf Left :} Metallicity of IC1459 within the 7.2'' pixel grid, unconstrained pixels are colored in gray, the average error on the value of each pixel is about 0.003. {\bf Right :}  radial profile of the metallicity of IC1459 (r$_e$ = 33''), error bars indicate the spread of the metallicity within each radial bin (error on each measurements is about 0.003). Previous measurements from \cite{annibali07,li07} are shown with different colors.}
\label{fig:meti}
\end{center}
\end{figure*}

\begin{figure*}[h!t]
\begin{center}
\includegraphics[width=8cm]{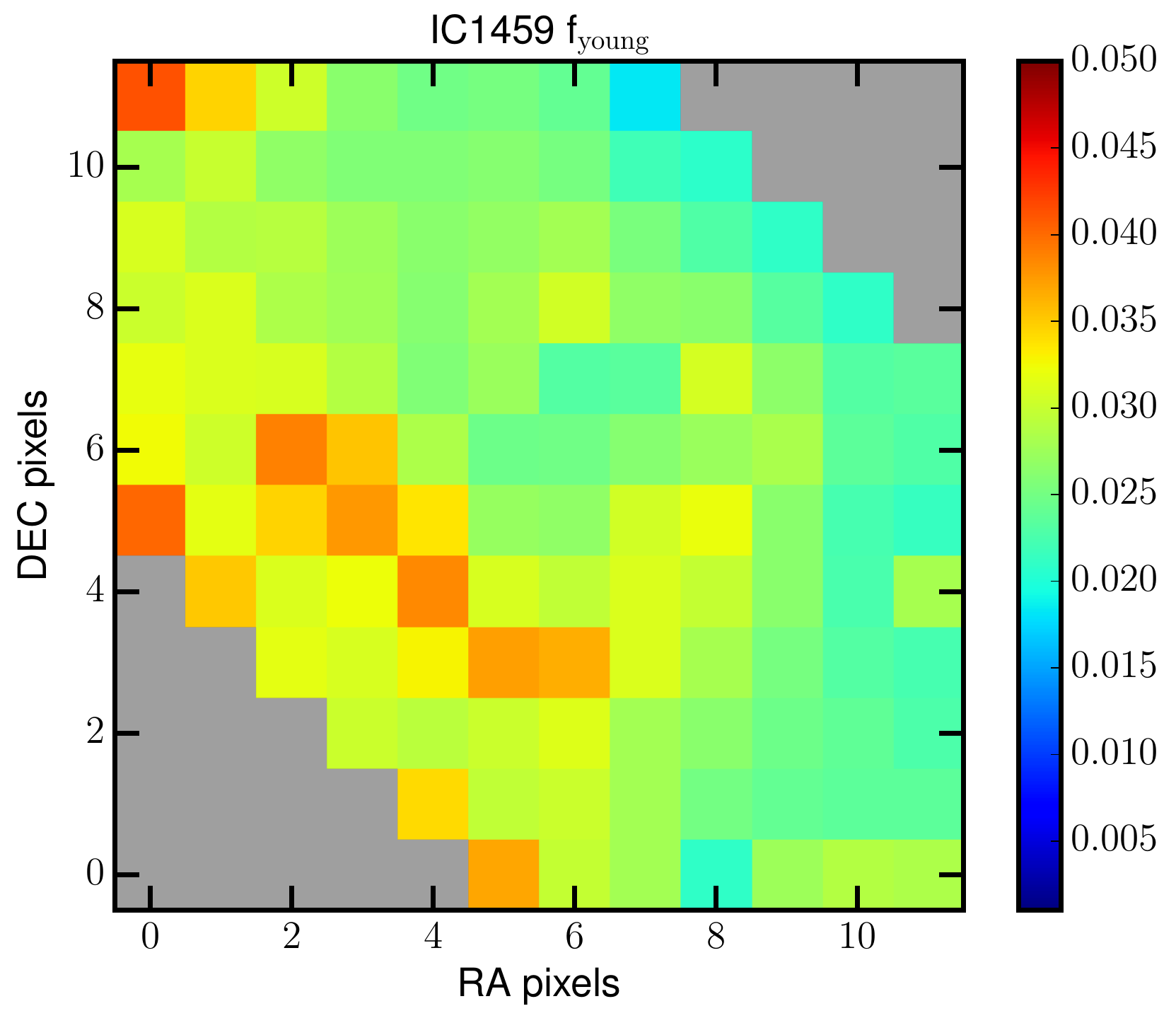}\hspace{0.5cm}\includegraphics[width=8.6cm]{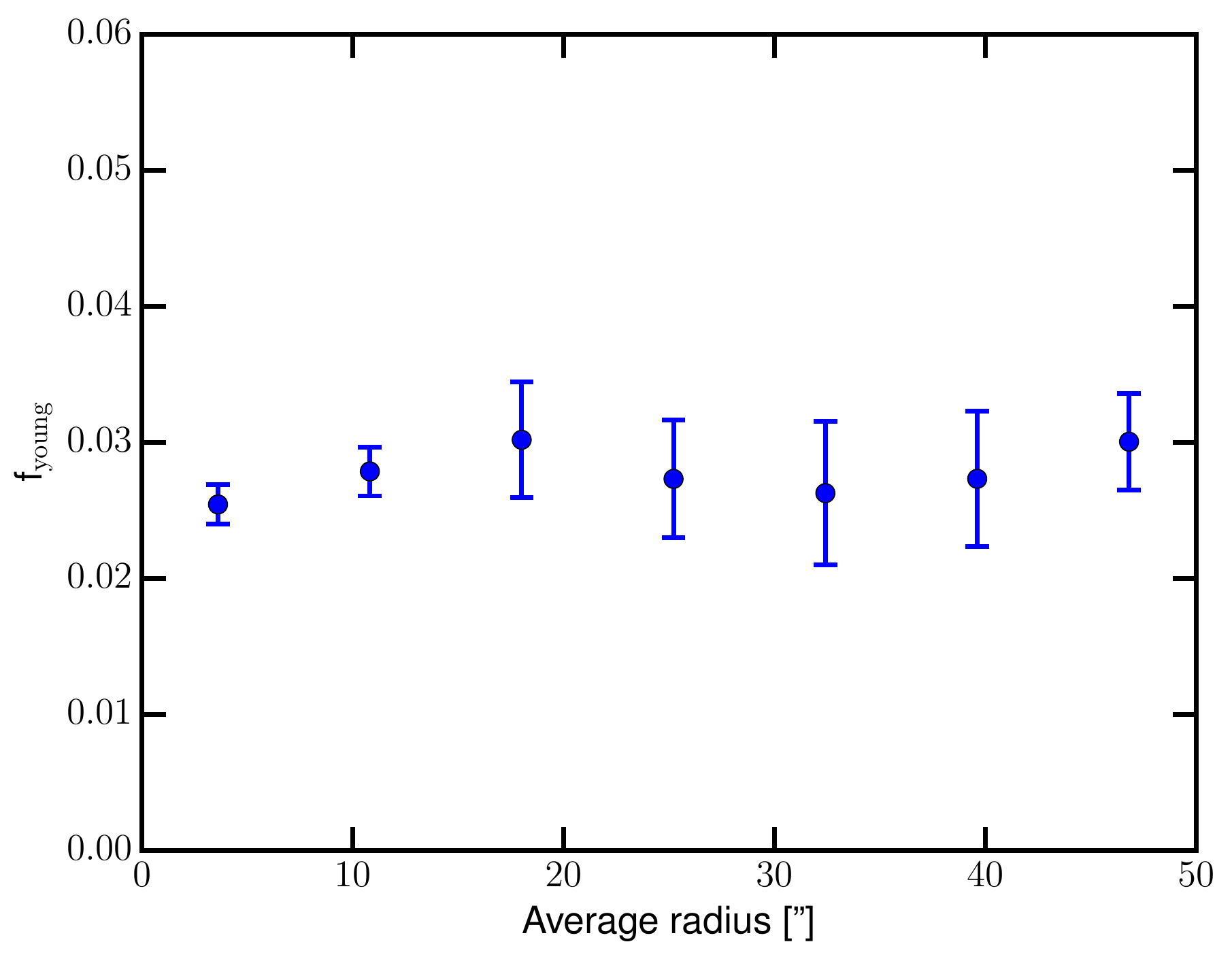}
\caption{{\bf Left :} fraction of young star in IC1459 within our grid of 7.2'' pixels, the average error on the derived values is about 0.005 (about 18\%). {\bf Right :} radial profile of the young stellar population fraction in IC1459, error bars indicate the spread of the young stellar population fraction within each radial bin (error on each measurements is about 0.005)}
\label{fig:fi}
\end{center}
\end{figure*}

In figure \ref{fig:fi}, the fraction of young stars fitted across IC1459 is on average about 3\% and peaks 
at about 4\% in the southwest strip corresponding to the lower metallicity and the PACS 160$\,\mu$m map arm. 
The central pixel of IC1459 has a young star fraction of about 2.5\%. 
The average error on the determination of the young star fraction is fairly substantial at about 0.5\%. 
The age and duration of the old stellar population formation is unconstrained and marginalized over the prior range, from 8 to 14 Gyr for its age and 0.5 to 3 Gyr for its e-folding time. \\
The average age of the young stellar population is about 0.4 Gyr with a typical error of 30 Myr, 
the duration (1 $\sigma$ of the Gaussian-shaped burst) of the recent burst of star formation 
is poorly constrained and only upper limits of about 80 Myr on average 
(95\% confidence level) are derived.\\ \cite{raimann01} modeled the age and metallicity of stars in IC1459
and found that 70\% of the stars on the outskirt (r > 20'')  and 90\% of the stars in the center (r < 4'') 
have an age of 10 Gyr with a solar metallicity. These results hint for a younger population on the outskirt
and a larger metallicity at the center, similarly to our results.
\cite{li07} found an average age of 4.9 $\pm$ 1.2 Gyr, that is hard to compare to
our model. \cite{serra10} found that IC1459 has a SSP-equivalent age 
of 3.5$^{+1.7}_{-0.4}$ Gyr using spectral line data from \cite{tal09}
and SSP models from \cite{thomas03}. This age estimate is biased towards the age of the youngest population 
\citep{serra07} and is more an indication of the age of young stars. Serra et al. 2010 gave
a rough estimate of 0.5 to 5 \% of young stellar population (in mass) formed between 300 Myr and 1 Gyr ago,
that estimate is in good agreement with our own.\\

\begin{figure*}[h!t]
\begin{center}
\includegraphics[width=8.5cm]{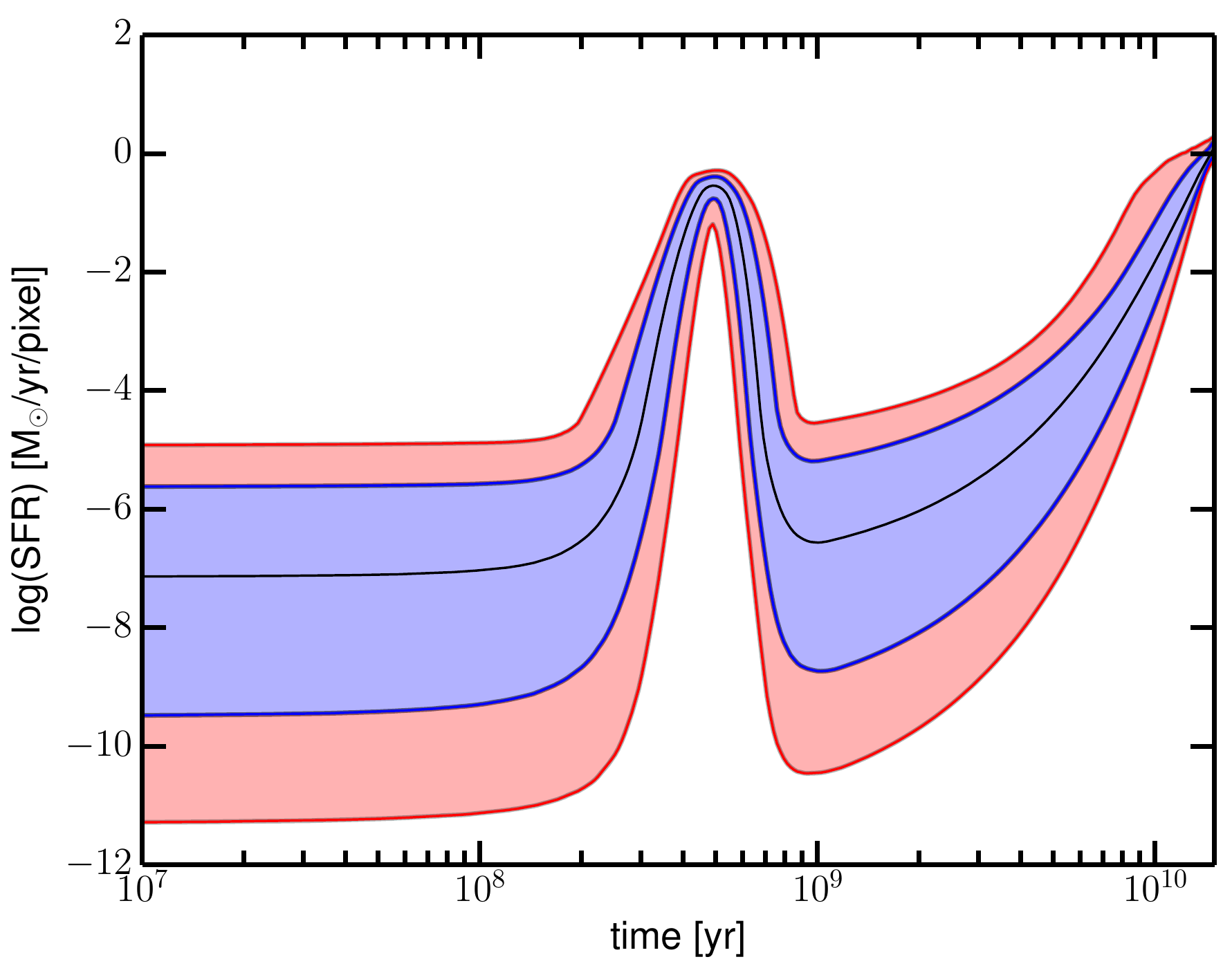}\hspace{0.5cm}\includegraphics[width=8cm]{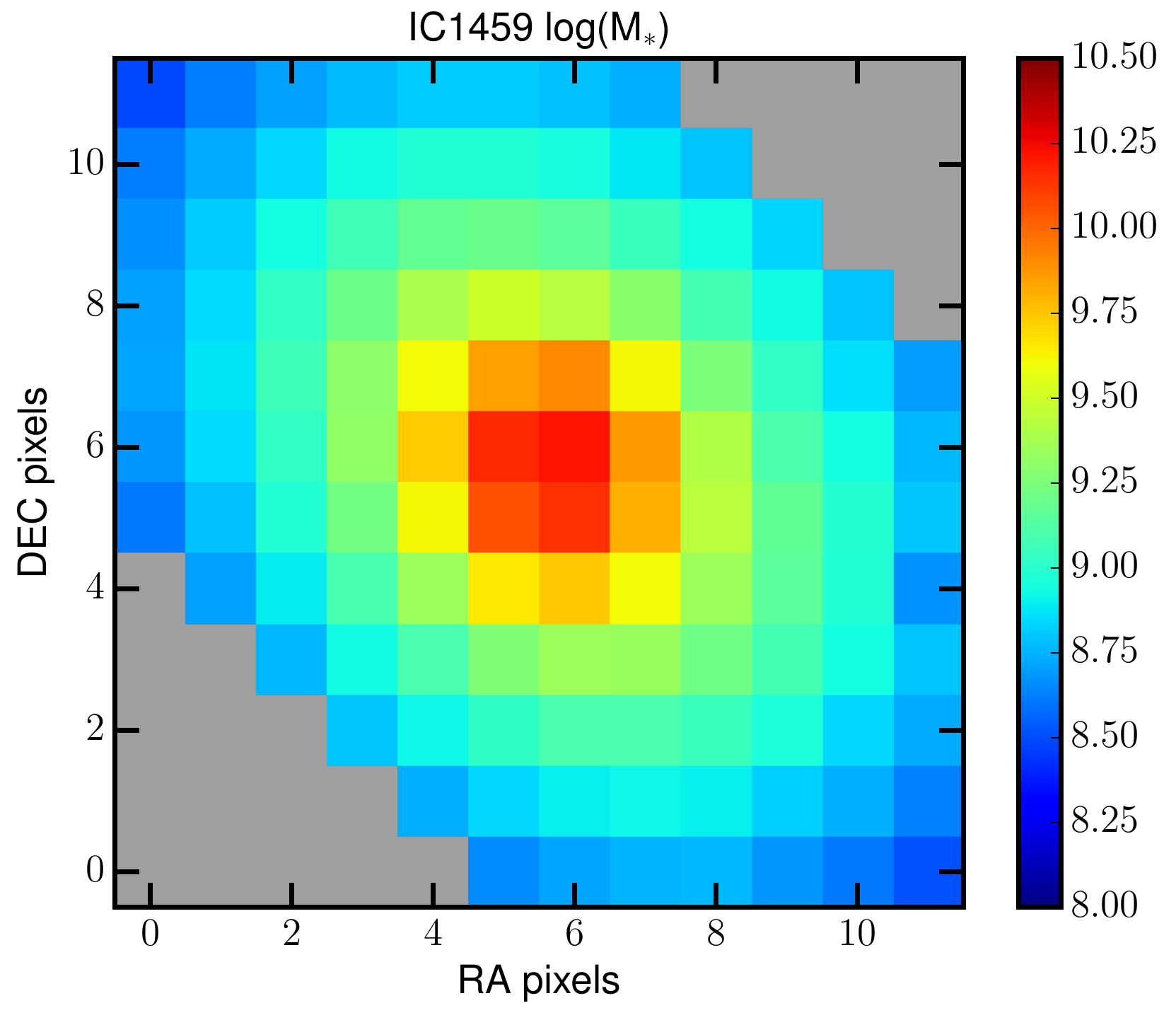}
\caption{{\bf Left :} SFR history of the pixel \#30 of IC1459, the blue contour represent the 68\% confidence level interval and the red contour represents the 95\% c.l. interval. {\bf Right :} stellar mass in IC1459 within our grid of 7.2'' pixels, the average error on the derived values is about 0.04 dex.}
\label{fig:sfri}
\end{center}
\end{figure*}

f$_{AGN}$ which represents the fractional contribution of a model of AGN to the total FIR luminosity, is constrained to be lower than 10\% at a 95\% confidence level where the FIR fluxes are quite large (i.e. PACS 160 $\mu$m map in figure \ref{fig:icgrid}). Elsewhere the constraints on f$_{AGN}$ are poorer and only a 20 to 25\% upper limit can be calculated.\\
\cite{dale02} model parameter $\alpha$ is measured to be 2.0 $\pm$ 0.1 in the central part of the galaxy (within a 15'' radius) and only constrained to be higher than 2.8 (95\% confidence level) elsewhere. 
The measured $\alpha$ translates into a measured dust temperature at the center of the galaxy (15'' radius) of 21.5 $\pm$ 0.9 K and an emissivity index of 2.2 $\pm$ 0.1.\\
The dust absorption in the V band shown in figure \ref{fig:avi} is constrained to be around 0.04 mag where PACS 160$\mu$m
is the largest, reaching about 0.06 at the center of the galaxy. This correlation shows the
consistency between the dust absorption near optical wavelengths and emission in the far-infrared.
Figure \ref{fig:mdi} shows a dust mass distribution that is, as expected, also correlated with the PACS 160$\mu$m image.
The dust mass radial profile slope is about -0.04 dex/arcsecond.

\begin{figure*}[h!t]
\begin{center}
\includegraphics[width=8cm]{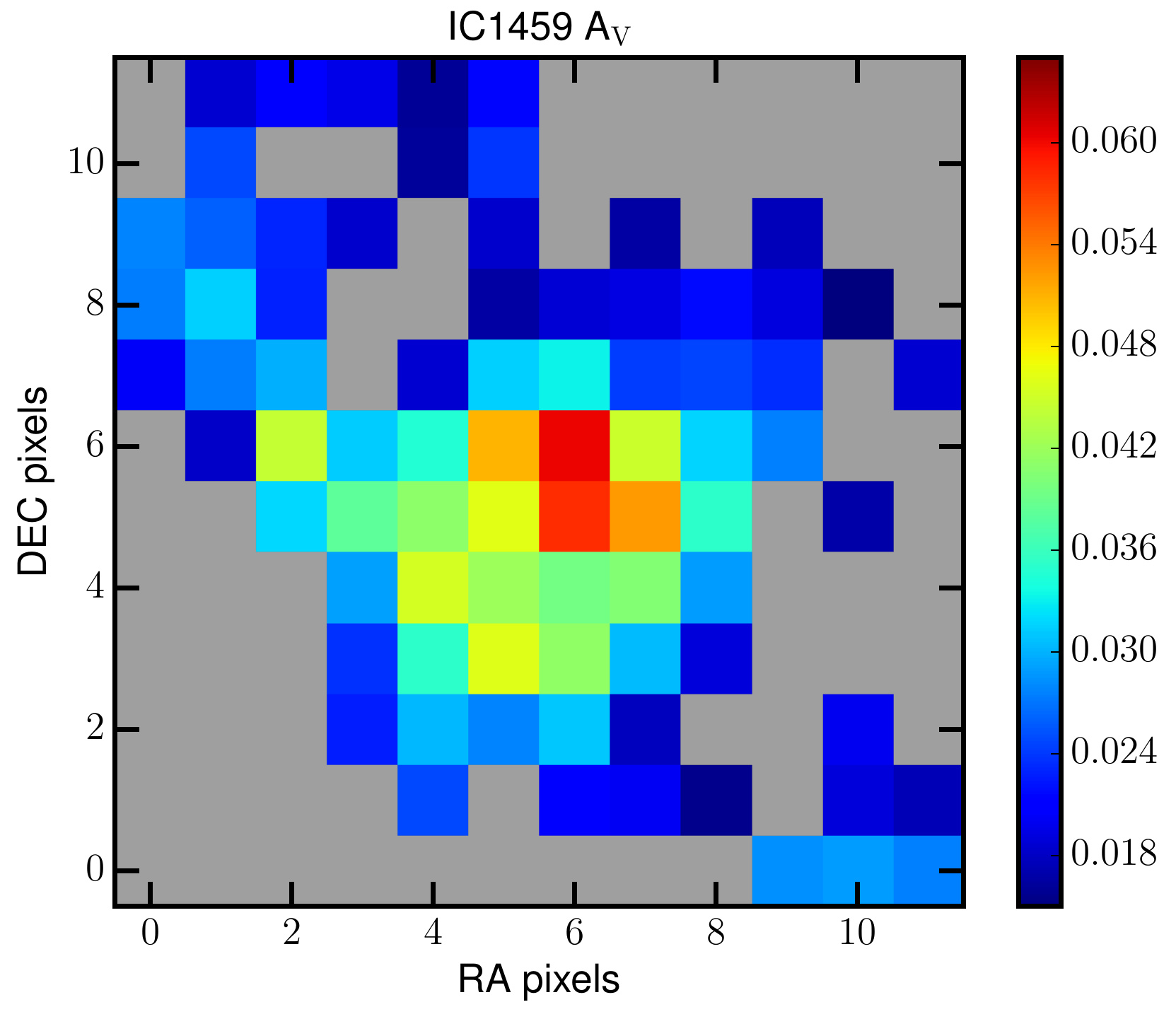}\hspace{0.5cm}\includegraphics[width=8.6cm]{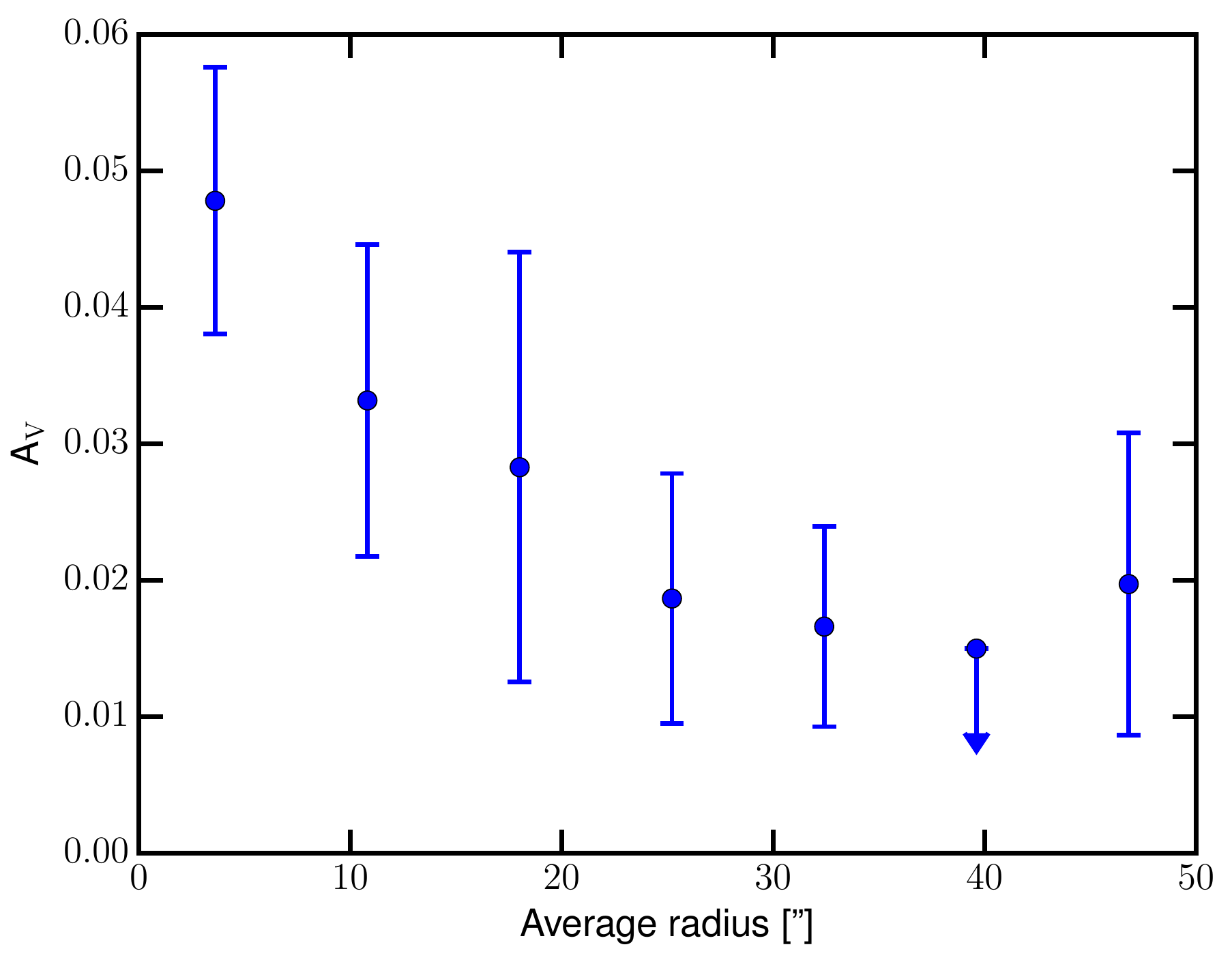}\\
\caption{{\bf Left :} dust absorption in the V band of IC1459 within the 7.2'' pixel grid, unconstrained pixels are colored in gray, the average error on the value of each pixel is about 0.004. {\bf Right :}  radial profile of the dust absorption of IC1459 (r$_e$ = 64''), error bars indicate the spread of the dust absorption within each radial bin.}
\label{fig:avi}
\end{center}
\end{figure*}
\begin{figure*}[h!t]
\begin{center}
\includegraphics[width=8cm]{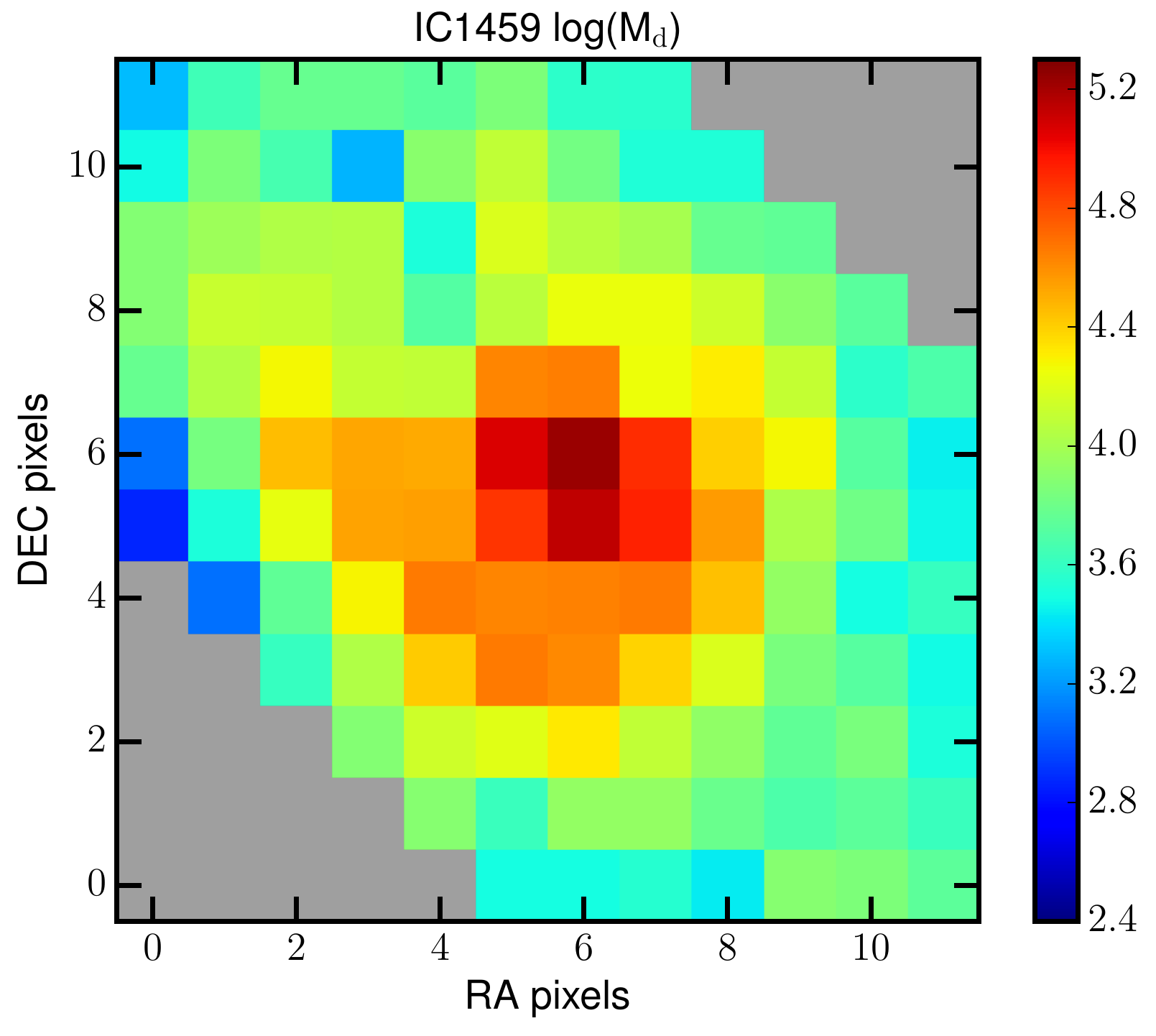}\hspace{0.5cm}\includegraphics[width=8.6cm]{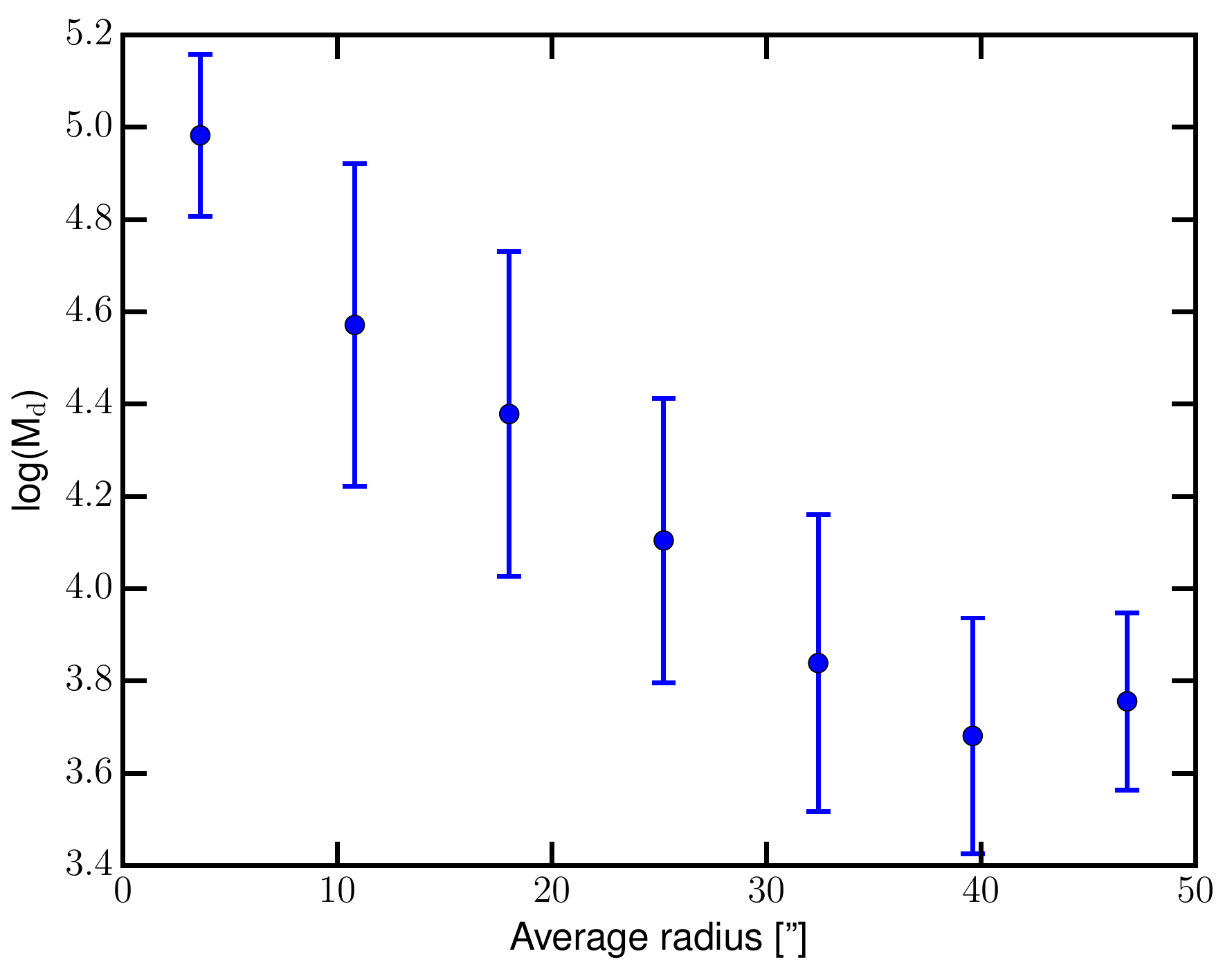}\\
\caption{{\bf Left :} dust mass of IC1459 within the 7.2'' pixel grid, unconstrained pixels are colored in gray, the average error on the value of each pixel is about 0.1. {\bf Right :}  radial profile of the dust mass of IC1459 (r$_e$ = 64''), error bars indicate the spread of the dust mass within the radial bin.}
\label{fig:mdi}
\end{center}
\end{figure*}

\section{Discussion}

NGC2768 and IC1459 are two elliptical galaxies with unusually large dust content as probed by 
dust absorption and emission. 
The model of stellar population synthesis, that we used in this work,
interprets this additional dust by a recent short burst of star formation that leads to an additional
3 to 4\% stellar mass.
The time scale of our bursts are short, less than about 100 Myr (1 $\sigma_{young}$
 at 95\% c.l., FWHM would be about 250 Myr). \cite{dimatteo07,dimatteo08} calculated from numerical simulation
of galaxy collisions, that most mergers produce a star formation burst of less than 500 Myr (the duration is calculated
during the period where the burst double the ``normal'' SFR).
\cite{dimatteo07,dimatteo08} also computed that most galaxy merger produce a maximum rate of about 25 M$_\odot$/yr with a median of 10 M$_\odot$/yr, the maximum SFR surface density being modeled at 10$^2$ to 10$^3$
M$_\odot$/yr/kpc$^2$.
Our fit on IC1459 SED returns a value of about 30 M$_\odot$/yr for the total maximum SFR, which is a bit high compared to \cite{dimatteo07} simulations. The highest SFR surface
density for IC1459 is 6 M$_\odot$/yr/kpc$^2$ and is more in line with these simulations.
NGC2768 total maximum SFR for the young star burst is about 10 M$_\odot$/yr and the maximum
SFR surface density is 1 M$_\odot$/yr/kpc$^2$. Both estimates of the maximum SFR and SFR density 
are assuming the 95\% confidence level upper limit on the duration of the recent SF burst, 
and therefore they constitute lower limits. IC1459 star formation
burst is a bit large compared to most mergers simulated in \cite{dimatteo07}, except for a retrograde
merger between giant ellipticals. \\
Another possibility would be that the gas producing the recent burst 
of star-formation has an internal origin contrary to the external
source of the merger hypothesis. \cite{gaspari15} explored how
the AGN interactions with the hot and cold gas could produce 
``in-situ condensation of the hot gas via radiative cooling and 
nonlinear thermal instability''. The condensation of the hot gas
could produce an episode of star formation and the AGN feedback
would disturb the gas morphology, producing filamentary gas structures
misaligned with the stellar distribution \citep{gaspari12}. 
Our SFH model is not a good representation of this model, 
but our SFH model tells us that the most recent significant burst 
of SF is about 0.5 Gyr and lasted less than 100 Myr. In the chaotic cold
accretion of \cite{gaspari12}, the gas cooling duration appears to be short
(\~ 100 Myr) and seems compatible with the short burst of star-formation
measured.
\\
\cite{kaviraj07} explored the UV color of a large catalog of elliptical galaxies, and found that most elliptical
with blue UV color (NUV-r < 5.5) had an episode of recent star formation mostly within the last Gyr which lead
to about an additional 1\% of young stars in mass. They also ran some merger models that can reproduce 
the NUV-r color distribution and give an average fraction of young star of about 1.4\% and an average age
of 0.6 Gyr. They also attempted to reproduce their data with a monolithic evolution model \citep{ferreras00}, where the gas
producing the recent star formation (RSF) episode is coming from the mass loss of old stellar population, but the
RSF can only produce 0.5\% of young star and does not reproduce the NUV-r color distribution.
NGC2768 and IC1459 have a blue NUV-r color (5.3 and 4.5 respectively) and our results match quantitatively
the results of \cite{kaviraj07} with a fraction of young stellar mass on the higher range.\\
With values between 0.025 and 0.05, the metallicities of our galaxies are consistent with most previous measurements 
for these galaxies and with the average metallicity of nearby elliptical galaxies \citep{annibali07,li07,silchenko06,howell05,denicolo05}.
The metallicity gradients are also consistent with previous measurements
for NGC2768 and IC1459 and with measurements of other early-type galaxies \citep{sanchezblazquez07,spolaor09,spolaor10,kuntschner10,rawle10,tortora10,labarbera11,koleva11}, with a medium to high negative gradient of -0.21 and -0.39 for NGC2768 and IC1459 respectively.\\
For instance, \cite{koleva11} found an average gradient for elliptical galaxies between 1 and 3 {\re}  of  -0.26 $\pm$ 0.08
using optical spectra and SSP models generated by Pegase.HR \citep{leborgne04}.\\
The metallicity gradients obtained for our two galaxies may be considered small for a quiescent evolution \citep{larson76,white80,carlberg84,kawata01,kobayashi04} where calculated gradients range between -0.5 to -1 for massive elliptical galaxies. The idea being that in the monolithic collapse, the outer part of the galaxy can not
withheld the gas as long as the central part of the galaxy. The central part is therefore fueled by more metal-rich gas able to create stars of higher metallicity in the later stages of the SF history. Recently however, some other studies \citep{pipino08,pipino10,tortora13} have found that the lower range of metallicity is compatible with quiescent evolution of elliptical galaxies. \cite{tortora13} explained the discrepancy by a high [$\alpha$/Fe] in the core of galaxies that shorten the SF and decreases the age gradients. In \cite{pipino10}, the average metallicity gradient is also quite large at -0.3 with a spread growing with the galaxy mass, while the SF efficiency seems to influence the gradient value for larger galaxies (larger efficiency returning more negative gradient).
On the other hand, metallicity gradients in post wet-merger galaxies are expected to be smaller at about a value of 
-0.2 to -0.3 \citep{kobayashi04,hopkins09}.\\
\cite{tortora10}, using a large sample of early-type galaxies, found that galaxies with the oldest stellar populations have the largest metallicity gradient. The relation has a large dispersion but it would indicate that IC1459 with a gradient of -0.39 has a stellar population age greater than 8 Gyr.\\
\cite{tortora10} found that color gradient in massive early type galaxies (M* > 10$^{11}$ M$_\odot$)
are quite shallow compared to gradient in late type galaxies or less massive early type.
They found that these color gradient are mostly due to metallicity gradient (age gradient being the second factor). Their minimum metallicity gradient reached -0.5 for these massive ETGs.

\section{Conclusions}

NGC2768 and IC1459 are two elliptical galaxies with an unusually large amount of dust,
seen in absorption and emission. In both cases, the dust spatial distribution does
not follow the stellar mass distribution. NGC2768 dust is distributed along the minor
axis in what is possibly a ring, IC1459 dust is distributed in some arms elongated
from the center of the galaxy either along the major axis or at a roughly 30 degree
angle from the major axis.\\
Using multi-wavelength coverage on NGC2768 and IC1459 from UV to FIR, we modeled the SED using
Maraston SPS and star formation history consisting of an exponentially decreasing old stellar
population and a Gaussian-shaped young stellar formation burst. From the parameters of our model, 
we derived that the dust distribution is associated with a larger fraction of young star ($\sim$ 3\%),
that were produced in a short burst ( < 100 Myr) about 0.5 Gyr ago. The dust mass
distributions follow the FIR maps for both galaxies, and are therefore correlated
with the fraction of young star. For both galaxies, the dust absorption location in the V band
is in good agreement with the dust emission. 
The age and duration of the older stellar population is not well constrained. 
These results are compatible with a recent merger but do not rule out an internal 
source for this recent burst of SF.
\\
The metallicity of both galaxies is well constrained within a 50'' radius around the center
and the gradients are found to be -0.21 and -0.39 respectively for NGC2768 and IC1459.
These gradients and metallicities are consistent with previous measurements and the typical
values for elliptical galaxies.

{\acknowledgements
This publication makes use of data products from the Two Micron All Sky Survey, which is a joint
project of the University of Massachusetts and the Infrared Processing and Analysis Center/California 
Institute of Technology, funded by the National Aeronautics and Space Administration and the
 National Science Foundation. This publication makes use of data from SDSS-III. Funding for SDSS-
III has been provided by the Alfred P. Sloan Foundation, the Participating Institutions, the National 
Science Foundation, and the U.S. Department of Energy Office of Science. The SDSS-III web site is 
http://www.sdss3.org/. SDSS-III is managed by the Astrophysical Research Consortium for the 
 Participating Institutions of the SDSS-III Collaboration including the University of Arizona, the 
Brazilian Participation Group, Brookhaven National Laboratory, University of Cambridge, Carnegie 
Mellon University, University of Florida, the French Participation Group, the German Participation 
Group, Harvard University, the Instituto de Astrofisica de Canarias, the Michigan State/Notre Dame/JINA 
Participation Group, Johns Hopkins University, Lawrence Berkeley National Laboratory, 
Max Planck Institute for Astrophysics, Max Planck Institute for Extraterrestrial Physics, New Mexico 
State University, New York University, Ohio State University, Pennsylvania State University, 
University of Portsmouth, Princeton University, the Spanish Participation Group, University of Tokyo, 
University of Utah, Vanderbilt University, University of Virginia, University of Washington,
 and Yale University. This work is based in part on observations made with the Spitzer Space Telescope, 
which is operated by the Jet Propulsion Laboratory, California Institute of Technology under 
a contract with NASA. This work is based in part on observations made with the NASA Galaxy Evolution 
Explorer. GALEX is operated for NASA by the California Institute of Technology under NASA contract .
This research has made use of the NASA/IPAC Extragalactic Database (NED) which is operated by the 
Jet Propulsion Laboratory, California Institute of Technology, under contract with the National 
Aeronautics and Space Administration. 
M.G. is supported by NASA through Einstein Postdoctoral Fellowship Award Number PF-160137 issued by the Chandra X-ray Observatory Center, which is operated by the SAO for and on behalf of NASA under contract NAS8-03060.
}

\bibliography{biblio}

\end{document}